\documentclass{article}
\usepackage{arxiv}
\usepackage[utf8]{inputenc}
\usepackage[T1]{fontenc}
\usepackage{graphicx}
\usepackage{amsmath,amssymb,amsfonts}
\usepackage{booktabs}
\usepackage{array}
\usepackage{tabularx}
\usepackage{placeins}
\usepackage{textcomp}
\usepackage{url}
\usepackage[numbers,sort&compress]{natbib}
\usepackage[colorlinks=true,linkcolor=blue,citecolor=blue,urlcolor=blue,hypertexnames=false]{hyperref}
\newcommand{\backmatter}{}
\newcommand{\bmhead}[1]{\paragraph*{#1}}
\graphicspath{{figures/}}
\expandafter\def\csname resultN001\endcsname{4,484}
\expandafter\def\csname resultN002\endcsname{2,591}
\expandafter\def\csname resultN003\endcsname{709}
\expandafter\def\csname resultN004\endcsname{655}
\expandafter\def\csname resultN005\endcsname{29 \% (8/28; 95 \% CI 13--49 \%)}
\expandafter\def\csname resultN006\endcsname{0 \% (0/28; 95 \% CI 0--12 \%)}
\expandafter\def\csname resultN007\endcsname{28 \% (7/25; 95 \% CI 12--49 \%)}
\expandafter\def\csname resultN008\endcsname{14 \% (4/28; 95 \% CI 4--33 \%)}
\expandafter\def\csname resultN009\endcsname{46 \% (13/28; 95 \% CI 28--66 \%)}
\expandafter\def\csname resultN010\endcsname{29 \% (8/28; 95 \% CI 13--49 \%)}
\expandafter\def\csname resultN012\endcsname{86 \% (24/28; 95 \% CI 67--96 \%)}
\expandafter\def\csname resultN014\endcsname{1.2\texttimes{} to 300,000\texttimes{}}
\expandafter\def\csname resultN015\endcsname{26 of 28}
\expandafter\def\csname resultN016\endcsname{134\texttimes{}}
\expandafter\def\csname resultN017\endcsname{18.6\texttimes{}}
\expandafter\def\csname resultN018\endcsname{3.47\texttimes{}}
\expandafter\def\csname resultN019\endcsname{18.0\texttimes{}}
\expandafter\def\csname resultN021\endcsname{2.32 million}
\expandafter\def\csname resultN022\endcsname{0.23 million}
\expandafter\def\csname resultN023\endcsname{1.92 million}
\expandafter\def\csname resultN024\endcsname{13.7 million}
\expandafter\def\csname resultN025\endcsname{1560 J}
\expandafter\def\csname resultN026\endcsname{102 J}
\expandafter\def\csname resultN028\endcsname{21 \% (6/28)}
\expandafter\def\csname resultN029\endcsname{23 \% (13/57)}
\expandafter\def\csname resultN030\endcsname{18 \% (10/57)}
\expandafter\def\csname resultN031\endcsname{19 \% (11/57)}
\expandafter\def\csname resultN032\endcsname{25 \% (14/57)}
\expandafter\def\csname resultN033\endcsname{14 \% (8/57)}
\expandafter\def\csname resultN034\endcsname{21 \% (6/28)}
\expandafter\def\csname resultN035\endcsname{\ensuremath{\kappa} = 0.35 (0.21--0.49)}
\expandafter\def\csname resultN036\endcsname{\ensuremath{\kappa} = 0.28 (0.12--0.43)}
\expandafter\def\csname resultN037\endcsname{\ensuremath{\kappa} = 0.66 (0.49--0.80)}
\expandafter\def\csname resultN038\endcsname{\ensuremath{\kappa} = 0.56 (0.42--0.70)}
\expandafter\def\csname resultN039\endcsname{0.62\texttimes{}}
\expandafter\def\csname resultN042\endcsname{0.47 s}
\expandafter\def\csname resultN044\endcsname{37.04 s}
\expandafter\def\csname resultN052\endcsname{3 of 5}
\expandafter\def\csname resultN053\endcsname{3 of 5}
\expandafter\def\csname resultN055\endcsname{79 \% (60 of 76)}
\expandafter\def\csname resultN062\endcsname{8.76 s}
\expandafter\def\csname resultN063\endcsname{1.52 s}
\expandafter\def\csname resultN064\endcsname{1.89 s}
\expandafter\def\csname resultN066\endcsname{9 \% (17/188; 95 \% CI 5--14 \%)}
\expandafter\def\csname resultN068\endcsname{4.9\texttimes{}}
\expandafter\def\csname resultN069\endcsname{10.8\texttimes{}}
\expandafter\def\csname resultN075\endcsname{6.5\texttimes{}}
\expandafter\def\csname resultN077\endcsname{57 of 100}
\expandafter\def\csname resultN079\endcsname{\ensuremath{\kappa} = 0.62 (0.42--0.80)}
\expandafter\def\csname resultN086\endcsname{57}
\expandafter\def\csname resultN087\endcsname{28}
\expandafter\def\csname resultN089\endcsname{7 of 25 (28 \%)}
\expandafter\def\csname resultN093\endcsname{1.74}
\expandafter\def\csname resultN095\endcsname{4.2\texttimes{}}
\expandafter\def\csname resultN096\endcsname{9.6\texttimes{}}
\expandafter\def\csname resultN101\endcsname{1.72\texttimes{}}
\expandafter\def\csname resultN102\endcsname{4 of 5}
\expandafter\def\csname resultN103\endcsname{46.6 million}
\expandafter\def\csname resultN105\endcsname{1.91\texttimes{}}
\expandafter\def\csname resultN106\endcsname{6.3\texttimes{} to 27.7\texttimes{}}
\expandafter\def\csname resultN108\endcsname{5 of 6}
\expandafter\def\csname resultN112\endcsname{2.0\texttimes{} to 6.5\texttimes{}}
\expandafter\def\csname resultN113\endcsname{15 of 28}
\expandafter\def\csname resultN114\endcsname{11 of 28}
\expandafter\def\csname resultN115\endcsname{7 of 8}
\expandafter\def\csname resultN119\endcsname{1 of 28}
\expandafter\def\csname resultN120\endcsname{0 of 28}
\expandafter\def\csname resultN122\endcsname{3.3\texttimes{} to 2,280\texttimes{}}
\expandafter\def\csname resultN123\endcsname{24.5\texttimes{}}
\expandafter\def\csname resultN128\endcsname{90}
\expandafter\def\csname resultN131\endcsname{27}
\expandafter\def\csname resultN132\endcsname{20 of 110}
\expandafter\def\csname resultN133\endcsname{90}
\expandafter\def\csname resultN138\endcsname{3,913\texttimes{}}
\expandafter\def\csname resultN139\endcsname{4.16 \%}
\expandafter\def\csname resultN140\endcsname{4.12 \%}
\expandafter\def\csname resultN144\endcsname{38 of 57 (67 \%)}
\expandafter\def\csname resultN146\endcsname{2}
\expandafter\def\csname resultN148\endcsname{168 H100-days}
\expandafter\def\csname resultN149\endcsname{16}
\expandafter\def\csname resultN151\endcsname{29}
\expandafter\def\csname resultN152\endcsname{15}
\expandafter\def\csname resultN155\endcsname{11 of 28}
\expandafter\def\csname resultN156\endcsname{10}
\expandafter\def\csname resultN158\endcsname{5}
\expandafter\def\csname resultN163\endcsname{0.76\texttimes{}}
\expandafter\def\csname resultN164\endcsname{15 of 25}
\expandafter\def\csname resultN165\endcsname{p = 0.42}
\expandafter\def\csname resultN167\endcsname{0.69\texttimes{}}
\expandafter\def\csname resultN172\endcsname{1.59\texttimes{}}
\expandafter\def\csname resultN173\endcsname{7 of 25}
\expandafter\def\csname resultN174\endcsname{p = 0.043}
\expandafter\def\csname resultN176\endcsname{1.47\texttimes{}}
\expandafter\def\csname resultN177\endcsname{8 of 25}
\expandafter\def\csname resultN178\endcsname{p = 0.11}
\expandafter\def\csname resultN204\endcsname{7}
\expandafter\def\csname resultN300\endcsname{seven}
\expandafter\def\csname resultN301\endcsname{six}
\expandafter\def\csname resultN303\endcsname{2,280--24,747\texttimes{}}
\expandafter\def\csname resultN304\endcsname{320--1,112\texttimes{}}
\expandafter\def\csname resultN305\endcsname{3.6\texttimes{}}
\expandafter\def\csname resultN306\endcsname{12--38}
\expandafter\def\csname resultN309\endcsname{4.7\texttimes{}}
\expandafter\def\csname resultN313\endcsname{1.6--1.7 \texttimes{} \ensuremath{10^{4}}}
\expandafter\def\csname resultN314\endcsname{1.85--1.96}
\expandafter\def\csname resultN315\endcsname{1.16\texttimes{}}
\expandafter\def\csname resultN316\endcsname{1.145\texttimes{}}
\expandafter\def\csname resultN319\endcsname{8.9 h}
\expandafter\def\csname resultN402\endcsname{4.2- to 20-fold}
\expandafter\def\csname resultN404\endcsname{37 million}
\expandafter\def\csname resultN405\endcsname{1.5\texttimes{} to 3.5\texttimes{}}
\expandafter\def\csname resultN406\endcsname{558}
\expandafter\def\csname resultN407\endcsname{0.59}
\expandafter\def\csname resultN408\endcsname{0.24}
\expandafter\def\csname resultN409\endcsname{0.19}
\expandafter\def\csname resultN410\endcsname{39 \% (11/28; 95 \% CI 22--59 \%)}
\expandafter\def\csname resultN413\endcsname{24 of 28}
\expandafter\def\csname resultN414\endcsname{10 of 24}
\expandafter\def\csname resultN415\endcsname{23 of 28}
\expandafter\def\csname resultN416\endcsname{5 of 23}
\expandafter\def\csname resultN423\endcsname{12 of 25}
\expandafter\def\csname resultN424\endcsname{1.54\texttimes{} to 7.65\texttimes{}}
\expandafter\def\csname resultN425\endcsname{13 of 25}
\expandafter\def\csname resultN428\endcsname{9 of 25}
\expandafter\def\csname resultN432\endcsname{4,635--4,741}
\expandafter\def\csname resultN433\endcsname{9,922}
\expandafter\def\csname resultN434\endcsname{8}
\expandafter\def\csname resultN435\endcsname{11}
\expandafter\def\csname resultN501\endcsname{5.3 million}
\expandafter\def\csname resultN502\endcsname{7.5 million}
\expandafter\def\csname resultN503\endcsname{24.7 million}
\expandafter\def\csname resultN504\endcsname{4.1 \%}
\expandafter\def\csname resultN505\endcsname{3.3 \%}
\expandafter\def\csname resultN506\endcsname{0.005}
\expandafter\def\csname resultN510\endcsname{0.61 (0.39--0.79)}
\expandafter\def\csname resultN511\endcsname{0.59 (0.47--0.71)}
\expandafter\def\csname resultN512\endcsname{0.23 (0.11--0.36)}
\expandafter\def\csname resultN513\endcsname{0.26 (0.11--0.41)}
\expandafter\def\csname resultN514\endcsname{0.47 (0.30--0.63)}
\expandafter\def\csname resultN515\endcsname{0.71 (0.58--0.83)}
\expandafter\def\csname resultN516\endcsname{0.48--0.56}
\expandafter\def\csname resultN517\endcsname{0.46--0.60}
\expandafter\def\csname resultN518\endcsname{0.16--0.25}
\expandafter\def\csname resultN519\endcsname{0.23--0.39}
\expandafter\def\csname resultN520\endcsname{0.42--0.56}
\expandafter\def\csname resultN521\endcsname{0.44--0.56}
\expandafter\def\csname resultN522\endcsname{\ensuremath{\kappa} = 0.16--0.35}
\expandafter\def\csname resultN523\endcsname{66 \% (21/32; 95 \% CI 47--81 \%)}
\expandafter\def\csname resultN524\endcsname{37 \% (14/38; 95 \% CI 22--54 \%)}
\expandafter\def\csname resultN525\endcsname{3 or 4}
\expandafter\def\csname resultN530\endcsname{20-fold}
\expandafter\def\csname resultN531\endcsname{5.5-fold}
\expandafter\def\csname resultN532\endcsname{1.0- to 1.4-fold}
\expandafter\def\csname resultN533\endcsname{153 of 250}
\expandafter\def\csname resultN534\endcsname{1.3 percentage points}
\expandafter\def\csname resultN535\endcsname{83 of 125}
\expandafter\def\csname resultN536\endcsname{78 of 125}
\expandafter\def\csname resultN537\endcsname{1.77\texttimes{}}
\expandafter\def\csname resultN538\endcsname{0.96\texttimes{}}

\newcommand{\N}[1]{\ifcsname result#1\endcsname\csname result#1\endcsname\else\errmessage{Unknown result token #1}\fi}
\newcommand{\ce}{\mathrm{CE}}
\title{How the Classical Baseline Shapes Machine-Learning Speed Claims in Topology Optimization}
\renewcommand{\shorttitle}{Classical baselines for learning-based topology optimization}
\author{
  Shaoliang Yang \\
  Santa Clara University \\
  Santa Clara, CA 95053, USA \\
  \texttt{syang11@scu.edu}
  \And
  Jun Wang\thanks{Corresponding author.} \\
  Santa Clara University \\
  Santa Clara, CA 95053, USA \\
  \texttt{jwang22@scu.edu}
}
\date{}
\begin{document}
\maketitle
\begin{abstract}Topology optimization places material in a design domain to optimize a structure's performance, such as its stiffness or heat conduction. Machine-learning methods often report large speed-ups over classical, non-learned optimization, but a slow classical code or a lower-quality learned design can exaggerate them. We re-ran released code for eight methods on one workstation and studied two in depth: Optimize Any Topology (OAT), a diffusion model that generates designs, and a graph-based neural optimizer. Replacing the classical code with an efficient implementation of the same algorithm cut these two methods' speed-ups by up to \N{N530}. We then timed the classical code until its design was at least as stiff as the learned one; for OAT, we picked this point after the runs, which gives the classical code its best case. Published timings imply a \N{N016} speed-up for OAT on a $256\times256$ grid; our matched single-design comparisons gave \N{N405}. At matched stiffness, the graph-based optimizer was slower than the classical code in all six two-dimensional problems. Of the six other methods, two kept their advantage, one shrank, one reversed, and two were not comparable. In an exploratory audit, language models judged that \N{N434} of \N{N087} speed comparisons in randomly sampled, openly accessible articles lacked evidence of equal design quality; our own review counted \N{N435}, so the count depends on who judges. We provide a comparison procedure and a minimum report: classical code, hardware, design quality, what each timing includes, and data-generation and training costs.\end{abstract}

\section{Introduction}
Topology optimization decides where to place material in a design domain so that a structure performs well, for example with the highest stiffness or the best heat conduction, under specified loads and constraints. Once trained, machine-learning methods can return such layouts quickly, and articles often report large speed-ups over classical optimization \citep{audP00550,audP00610,audP00827}; in a random sample of articles that we audit below, \N{N015} speed claims gave a numeric factor, with median \N{N123} (Appendix~\ref{app:published}). A speed-up, however, compares two timings that may not cover the same work: a learned design may still need a mechanical evaluation, constraint checks, or refinement before it can be used. A fair comparison therefore measures the time each method needs to reach a stated design quality, such as stiffness under the specified loads, and states what each timer includes, which we call the timing boundary.

The classical implementation matters even when the optimization algorithm stays fixed. For example, the solid isotropic material with penalization (SIMP) method repeatedly computes how a candidate layout deforms under the loads, then updates the material distribution \citep{andreassen2011}. This evaluation solves the equations of mechanical equilibrium, and a faster solver for those equations can shorten the runtime while the optimization algorithm stays the same \citep{amir2014}. Parallel execution on many processor cores or a graphics processing unit (GPU) can shorten the runtime further \citep{traff2023,aage2017}. A learning-based method compared with a slow classical solver, or timed on a different device, can therefore receive credit for an implementation advantage. Reviews of the field already ask for comparisons at equal design quality, and audits in other areas of machine learning found that reported gains shrank under corrected protocols or lost to simple baselines (Sect.~\ref{sec:related}).

How do classical implementation and matched design quality change a learning-based method's measured speed advantage?

We answer this question by testing released code for eight methods on one workstation. We selected two methods for detailed tests before any timed run: Optimize Any Topology (OAT), a diffusion model that generates a design by sampling \citep{nobari2025}, and HGTO, a graph-based method that optimizes a neural-network representation of the design \citep{liu2026}. Both release runnable code and report timings against their own classical code, allowing controlled re-testing. Neither belongs to the sample of published articles that we audit below; the other six methods come from that sample. Each of the eight articles compares the speed of its learning-based method with a classical method, as a stated factor or through reported or projected run times. Both detailed re-tests concern stiffness problems, scored by compliance (lower is stiffer); the other six methods and the article audit also cover thermal design, while fluid, acoustic, photonic, and electromagnetic topology optimization lie outside this study.

The two detailed re-tests use a ladder of four speed-ups, where a speed-up is the classical time divided by the learning-based method's time (Fig.~\ref{fig:procedure}). The first rung is implied by the timings published in the method's article. The second re-runs the released learning-based and classical codes on our workstation. The third replaces the classical code with an efficient implementation of the same algorithm, and the fourth times that implementation only until its design matches the learned design's quality, as scored by the code released with the learning-based method. For OAT on a $256\times256$ grid, the speed-up falls along these rungs from \N{N016} to 78.6$\times$, \N{N017}, and finally \N{N018}. Here the last rung uses a matching point found after the runs, which gives the classical code its best case.

To see whether published comparisons report these conditions, we also audited \N{N086} articles from a random sample; \N{N087} of them make a speed claim. Two language models read the openly accessible full texts and checked four rules: whether the compared designs have matched quality, whether the classical baseline is efficient, whether data-generation and training cost is reported, and whether the returned design was verified. A language model resolved their disagreements, and we then reviewed the verdicts ourselves while knowing what the models had decided. These judgments describe the sampled articles and depend on who judges. Seven of the speed claims linked public code; we call them the census and tried to re-run all seven. We ran code for six, the six methods above; the seventh, the predictor of \citet{halle2021}, stopped before timing at a reconstruction check fixed in advance. The census is complete for this sample but does not represent released methods in general, and each census test uses its article's own comparison and quality measure (Sect.~\ref{sec:census-procedure}).

The efficient classical implementation alone cut the same-machine speed-ups of the two in-depth methods by up to \N{N530}, the largest cut being HGTO's on its finest finite-element mesh. Matching design quality then left OAT \N{N405} faster at {\unboldmath$256^2$}, depending on the quality rule and problem set, and it made HGTO slower than the classical implementation in all six of its tested problems. In the census, two methods kept their advantage, one advantage shrank, one reversed, and two could not be compared, so the results do not support a blanket verdict on learning-based optimization. In the article audit, language models judged that \N{N434} of \N{N087} speed comparisons lacked evidence of equal design quality, and our own review counted \N{N435}.

The study makes three contributions:
\begin{enumerate}
\item A comparison ladder (Fig.~\ref{fig:procedure}) and minimum report (Sect.~\ref{sec:minimum}) connect a published speed ratio to execution conditions, design quality, and the offline cost of data generation and training.
\item Same-machine tests of released code for eight methods, including two detailed re-tests, measure how classical implementation and evaluated design quality change speed comparisons.
\item An exploratory, fixed-rule article audit records reporting gaps and how verdicts vary between judges, and its public-code census covers every sampled speed claim that linked public code.
\end{enumerate}
Sect.~\ref{sec:related} places the study in related work, and Sect.~\ref{sec:methods} describes the methods. Sect.~\ref{sec:results} reports these outcomes, and Sect.~\ref{sec:discussion} explains their scope and the information needed to reconstruct a speed claim.

\section{Related work}\label{sec:related}
Three lines of work bear on speed claims for learning-based topology optimization: evaluations of the learning-based methods, efficient classical implementations, and audits of evaluation practice elsewhere in machine learning.

\subsection{Learning-based topology optimization and its evaluation}
Reviews of learning-based methods for topology optimization describe neural predictors, generators, and surrogates \citep{shin2023,regenwetter2022}, and they ask for comparisons of designs of equal quality, with explicit costs and sound benchmarks \citep{woldseth2022,sigmund2022}. Controlled studies add that the outcome can depend on the problem. For example, optimizing a neural network that represents the design (neural reparameterization) may help find useful local optima, but its extra non-convexity may delay convergence \citep{herrmann2024,sanu2025}. Learning-based methods also carry an offline cost, the time to generate training data and to train, which has been accounted for in detail for one method \citep{hamann2026}. \citet{habibi2026} weighed the data-generation, training, and refinement costs of three inverse-design model families against running topology optimization alone. Together, these studies suggest testing the classical implementation and the quality threshold together, rather than treating the ratio of inference times as a complete performance measure.

\subsection{Efficient classical implementations}\label{sec:related-classical}
Efficient classical topology optimization has its own literature, and the efficient implementations in this study use only published components. The SIMP code with a density filter and optimality-criteria update appeared in 2011 \citep{andreassen2011}, multigrid-preconditioned conjugate gradients in 2014 \citep{amir2014}, the sparse direct solvers by 2005 \citep{schenk2004,li2005}, and public GPU topology-optimization codes in 2023 \citep{traff2023}. These dates matter because a method's authors could compare only against components published before their method; Sect.~\ref{sec:classical} checks this order for each comparison.

\subsection{Evaluation practice elsewhere in machine learning}
Other areas of machine learning show the same evaluation problem. In neural recommendation, most studied methods could not be reproduced or were often beaten by simple heuristics \citep{ferraridacrema2019}. Reported gains in deep metric learning shrank to marginal ones under a corrected protocol \citep{musgrave2020}, while data leakage produced overoptimistic conclusions across many fields \citep{kapoor2023}. The closest precedent in scientific machine learning audited learned solvers for the partial differential equations of fluid flow and found many comparisons against weak baselines \citep{mcgreivy2024}. Because that audit did not cover solid mechanics, it cannot show how baseline choices affect topology-optimization claims.

\subsection{Position of this study}
\citet{woldseth2022} called for comparable objectives and cost accounting in neural topology optimization, and \citet{sigmund2022} emphasized benchmarking. We operationalize these concerns as fixed rules for a defined sample and add same-machine implementation tests. The fluid-equation audit \citep{mcgreivy2024} motivated the rules. Table~\ref{tab:related} places this study among these reviews and evaluations.

\begin{table}[htbp]\caption{Related reviews and evaluations, and the contribution of this study.}\label{tab:related}\centering\scriptsize
\begin{tabularx}{\linewidth}{@{}>{\raggedright\arraybackslash}p{0.19\linewidth}>{\raggedright\arraybackslash}p{0.16\linewidth}>{\raggedright\arraybackslash}p{0.18\linewidth}>{\raggedright\arraybackslash}X@{}}\toprule Study&Domain&Design&Main finding or recommendation\\\midrule
\citet{woldseth2022}&Neural topology optimization&Critical review&Asks for comparisons at equal volume and quality, and for training cost in any break-even claim\\
\citet{sigmund2022}&Topology optimization&Recommendations&Recommends fair evaluation and benchmarking practice\\
\citet{shin2023}&Topology optimization&Review&Surveys machine-learning-based topology optimization\\
\citet{regenwetter2022}&Engineering design&Review&Surveys deep generative models in design, including structural optimization\\
\citet{herrmann2024}&Topology optimization&Controlled study&Neural-network discretizations can find better local optima under certain conditions\\
\citet{sanu2025}&Topology optimization&Controlled study&Neural networks add non-convexity, which can delay convergence in convex problems\\
\citet{hamann2026}&Topology optimization&Cost accounting, one method&Data generation and training pay off only after a very large number of applications\\
\citet{habibi2026}&Topology optimization&Cost--benefit study&Compares data-generation, training, and refinement costs of three inverse-design model families with running the optimizer alone\\
\citet{mcgreivy2024}&Fluid-equation solvers&Systematic review&Weak baselines in \N{N055} of included articles\\
\citet{ferraridacrema2019}&Recommender systems&Reproduction&Of 18 neural methods, 7 were reproducible with reasonable effort; 6 of these were often beaten by simple heuristics\\
\citet{musgrave2020}&Deep metric learning&Re-evaluation&Under a corrected protocol, improvements over time were marginal at best\\
\citet{kapoor2023}&Machine-learning-based science&Survey of leakage&Data leakage in 17 fields affected 294 papers, in some cases with overoptimistic conclusions\\
\citet{kapoor2024reforms}&Machine-learning-based science&Consensus reporting recommendations&32 items cover study goals, computational reproducibility, evaluation, and the scope of claims\\
\midrule
This study&Topology optimization&Fixed-rule audit of \N{N086} sampled articles; re-runs of eight released methods on one workstation&Unmatched comparisons in \N{N010} of speed claims as judged by language models (\N{N435}/\N{N087} in the authors' review); an efficient classical implementation cut the same-machine speed-ups of the two in-depth re-tests by up to \N{N530}\\
\bottomrule
\end{tabularx}\end{table}

\section{Methods}\label{sec:methods}
The study combines two controlled re-tests, a census of every public-code speed claim in the audited sample, and an exploratory audit of published comparisons. The re-runs test what changes when the classical implementation and quality threshold are controlled. The audit records what accessible article texts report. We describe the formulation and timing procedure first, then the re-tests, census, and article coding; Table~\ref{tab:overview} summarizes the four parts and where their results appear. Steps called pre-registered were fixed in writing before the coding or runs they govern, in hashed plans or dated pre-run log rows. The hashes are held in our study records without an external registry timestamp. Later changes are labeled post hoc and logged with date and reason (Online Resource 1).

\begin{table}[htbp]\caption{Parts of the study. The re-tests use the comparison ladder of Fig.~\ref{fig:procedure}: $S_{\rm pub}$ is implied by an article's published times, $S_1$ re-runs the released codes on our workstation, $S_2$ gives the classical code an efficient state solve, and $S_3$ times it until its design reaches the learned design's quality. The census levels L0--L2 apply the same steps to each article's own comparison. R1--R4 are the four audit rules (Sect.~\ref{sec:rules-summary}): R1 matched quality or time, R2 efficient baseline, R3 full cost, and R4 verified output. GPU, graphics processing unit.}\label{tab:overview}\centering\scriptsize
\begin{tabularx}{\linewidth}{@{}>{\raggedright\arraybackslash}p{0.13\linewidth}>{\raggedright\arraybackslash}X>{\raggedright\arraybackslash}p{0.23\linewidth}>{\raggedright\arraybackslash}p{0.13\linewidth}>{\raggedright\arraybackslash}p{0.11\linewidth}@{}}\toprule Part&What is tested&Quality criterion&Ladder levels&Results\\\midrule
OAT re-test (Sect.~\ref{sec:oat-rules})&A diffusion model against our efficient classical code on five test problems, fixed in advance, at $64^2$ and $256^2$; classical runs of 5 to 150 iterations&Compliance error from OAT's evaluator; a volume condition, added after the runs, also limits material&$S_{\rm pub}$, $S_1$, $S_2$, $S_3$&Sects.~\ref{sec:implementation-results} and \ref{sec:oat-results}\\
HGTO re-test (Sect.~\ref{sec:hgto-procedure})&A graph-based method against the authors' classical code and our GPU port of it on six linear two-dimensional problems&Time to convergence, and time to reach HGTO's final compliance from its independent evaluator&$S_{\rm pub}$, $S_1$, $S_2$, $S_3$&Sects.~\ref{sec:implementation-results} and \ref{sec:hgto-results}\\
Census (Sect.~\ref{sec:census-procedure})&The six sampled speed claims whose public code ran, each in the article's own comparison&Each article's own quality measure (Table~\ref{tab:census})&L0 (code runs), L1 (as $S_1$), L2 (as $S_2$--$S_3$)&Sect.~\ref{sec:census-results}\\
Article audit (Sects.~\ref{sec:search}--\ref{sec:coding})&\N{N086} sampled articles, \N{N087} with speed claims, coded against rules R1--R4 by two language-model passes with model adjudication, then reviewed by the authors&Rule R1: matched quality or time&--&Sects.~\ref{sec:audit-results}--\ref{sec:agreement}\\\bottomrule
\end{tabularx}\end{table}

\subsection{Classical formulation and efficient implementation}\label{sec:classical}
The controlled tests change how mechanical equilibrium is solved while holding the optimization algorithm fixed. Compliance measures deformation under load; lower compliance means a stiffer design for the same loading and constraints. A finite-element evaluation computes this response by dividing the structure into small elements. The formulation below specifies the objective, equilibrium, and physical-volume conditions that the implementations share:
\begin{equation}
\min_{\boldsymbol\rho}\ c=\mathbf f^{\mathsf T}\mathbf u
\quad\text{s.t.}\quad
\mathbf K(\tilde{\boldsymbol\rho})\mathbf u=\mathbf f,\quad
\sum_e\tilde\rho_e v_e\le V^*,\quad 0\le\rho_e\le1 .
\label{eq:form}
\end{equation}
Here $\rho_e$ is the design density of element $e$, $\tilde\rho_e$ its filtered density, $\mathbf f$ the load vector, $\mathbf u$ the displacements, $\mathbf K$ the stiffness matrix, $v_e$ the element volume, and $V^*$ the volume limit. Within each controlled comparison, three parts of the algorithm stay fixed: the material interpolation, which turns densities into stiffness; the density filter, which smooths the densities; and the optimality-criteria (OC) update, which changes them at each iteration \citep{andreassen2011}. The optimization algorithm stays fixed; its implementation, including the equilibrium (state) solver and the device, changes. Appendix~\ref{app:solver} gives the interpolation, filter, update, and material parameters.

Our efficient implementation uses preconditioned conjugate gradient (CG), an iterative equation solver, on a graphics processing unit (GPU) \citep{amir2014}. The sparse direct solver PARDISO provides an alternative on the central processing unit (CPU) \citep{schenk2004}. Before timing, both reproduced the compliance and physical (filtered) density computed by SuperLU, the reference sparse direct solver, within pre-registered tolerances \citep{li2005}. Every component of the efficient implementations was published before the learning-based methods whose comparisons it changed (Sect.~\ref{sec:related-classical}); public GPU topology-optimization codes appeared in 2023 \citep{traff2023}, before OAT, HGTO, and VoroTO, the three comparisons that used a GPU. The two compared census methods published before 2023, in 2019 and 2022, kept their own-code comparison at L2 (Table~\ref{tab:census}). Appendix~\ref{app:solver} gives the solver details, and Online Resource 2 records the per-case checks.

\subsection{Timing and energy}
All same-machine ratios use an Intel Core i5-14600KF workstation with 95.8 gibibytes (GiB) of random-access memory (RAM) and an NVIDIA RTX 5090 under the CUDA 12.8 GPU computing platform (CUDA 13.0 for OAT's classical code, which uses the CuPy GPU array library), running Windows 11. Timed processes were pinned to the performance cores except one census claim run in a container (Sect.~\ref{sec:limitations}). Warm-ups preceded timing, and GPU synchronization enclosed each timed interval. Repetition schedules and variability checks are reported in Appendix~\ref{app:timing}.

Each ratio depends on its timing boundary, that is, on what the clock includes. Our classical times in the OAT comparison include filter and solver setup, whereas the times of OAT's released classical code exclude its setup, which favors that code. OAT's inference time covers one sample, without model loading or batch assembly. Online Resource 2 lists the hardware and timing boundaries.

\begin{figure}[htbp]\centering\includegraphics[width=\linewidth]{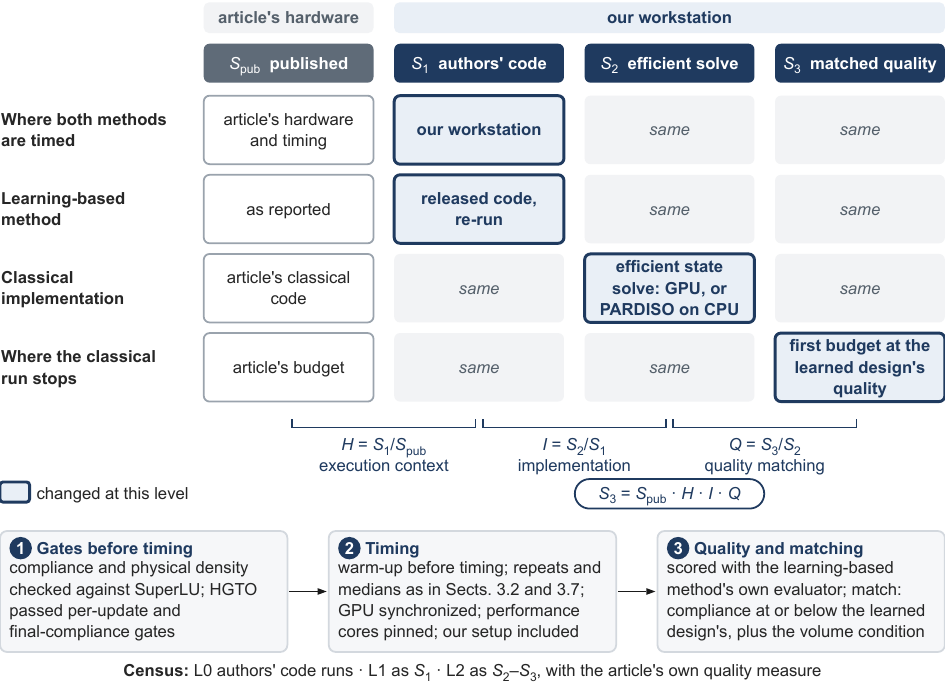}
\caption{Comparison levels and re-test procedure. Each row is one factor of the comparison; a highlighted cell marks the level at which it changes, and ``same'' repeats the value to its left. $S_{\rm pub}$ is the ratio of the times an article reports on its own hardware; $S_1$ re-runs the article's classical code and the learning-based method on our workstation; $S_2$ keeps the classical algorithm and budget with an efficient state solve; $S_3$ compares the time needed for that implementation to reach the learned design's evaluated quality, with matches for OAT (Optimize Any Topology, a diffusion model) identified after the runs. GPU/CPU, graphics/central processing unit; PARDISO and SuperLU are sparse direct equation solvers. The level-to-level ratios $H$, $I$, and $Q$ form the identity in Eq.~(\ref{eq:decomp}). Correctness gates precede timing, quality is scored with the learning-based method's own evaluator, and census levels L0--L2 adapt these steps to the article-specific quality criteria of Sect.~\ref{sec:census-procedure}}\label{fig:procedure}\end{figure}

The comparison ladder (Fig.~\ref{fig:procedure}) separates three influences on a speed-up: the execution context, the classical implementation, and the design quality. Each rung $S$ is a speed-up, the classical time divided by the learning-based method's time, so $S>1$ favors the learning-based method. $S_{\rm pub}$ is implied by the times an article reports on its own hardware; it gives context but is not a same-machine measurement. On our workstation, $S_1$ re-runs the article's classical code. $S_2$ uses our efficient implementation of the same algorithm for the same number of iterations. $S_3$ then times that implementation until it reaches the learned design's quality. Comparing $S_1$ with $S_2$ isolates the implementation, and comparing $S_2$ with $S_3$ isolates the quality threshold.

For OAT at $256^2$, the ladder falls from \N{N016} (published) to 78.6$\times$ (both released codes on our workstation), \N{N017} (our efficient GPU implementation at the same 150-iteration budget), and \N{N018} (our implementation timed to OAT's quality under the volume-condition rule of Sect.~\ref{sec:oat-rules}, which also limits material).

We chose the GPU implementation as the primary OAT baseline after the runs and also report the CPU alternative. Appendix~\ref{app:timing} gives its timing and the energy-measurement details.

\subsection{Re-test and census procedures}
The two in-depth methods were selected for reconstructable comparisons, not sampled to represent the literature. Both release runnable code, report timings against their own classical code, and test linear two-dimensional compliance problems. OAT also releases trained checkpoints. Neither belongs to the audited article sample, so their results enter no audit rate. Both were chosen before any timed run.

A third in-depth re-test, of the predictor of \citet{halle2021}, stopped before timing: our port of its classical code did not reproduce the article's stored results closely enough to pass the pre-registered check. This limits our reconstruction and does not assess the article's results (Appendix~\ref{app:selection}; Online Resource 5). The predictor is therefore excluded from the eight released-code tests. It is also the census's seventh claim (Sect.~\ref{sec:census-procedure}).

\subsubsection{OAT: time to matching design quality}\label{sec:oat-rules}
OAT generates a material layout by sampling its released diffusion model. We score every design with OAT's own evaluator, which reports the compliance error (CE): the percentage difference between a design's compliance and that of the reference design stored with each problem in OAT's dataset, so a lower CE means a stiffer design. Appendix~\ref{app:comparison} gives the sampling settings and the classical iteration budgets.

We report two quality rules for OAT. Under the pre-registered rule, a classical design matches when its compliance error is no greater than OAT's: $\ce_{\rm ours}\le\ce_{\rm OAT}$. This compares stiffness alone.

Before scoring, each design is thresholded: every intermediate density is rounded to solid or void. This can add material and make the design stiffer, reducing its compliance error without better optimization. We therefore added a volume condition after the runs. A match must satisfy the compliance-error rule and use no more material than the larger of the target volume and OAT's own volume. We use this stricter rule for the primary comparison and report the pre-registered rule alongside it. The added condition can delay or remove a match.

We ran the classical optimizer afresh for each budget from 5 to 150 iterations. The matched time is the run time of the smallest budget that meets the rule. Because this budget was identified after all runs, and OAT's evaluator was not timed, the comparison does not measure an online stopping rule. It shows how fast the classical code can reach OAT's quality, not how a user would know when to stop. Appendix~\ref{app:oat-extension} describes two post hoc additions: an extension to 25 problems per grid size, and a best-of-five sensitivity that keeps the lowest-error of five OAT samples.

\subsubsection{HGTO: convergence and matched quality}\label{sec:hgto-procedure}
HGTO optimizes a graph-based neural representation of the design \citep{liu2026}. We re-ran six linear two-dimensional problems and compared HGTO with the authors' classical SIMP code, which uses the OC update (SIMP--OC), and with our GPU port of the same algorithm. The port kept the update and stopping rules and passed the pre-timing checks. Appendix~\ref{app:comparison} gives the cases and algorithm settings.

The primary outcome is the time to convergence under the termination rule released with the code. The secondary outcome is the time the classical port needs to reach HGTO's final compliance and volume, scored with HGTO's independent evaluator \citep{gustafsson2020}. The first measures reaching a stopping condition; the second measures reaching the learning-based method's design quality. Cases with large deformations were not tested.

\subsubsection{Public-code census: the article's comparison}\label{sec:census-procedure}
The census covers every speed claim in the audited sample whose article links public code. Its candidates and the rules for repairs needed to run released code were fixed before any timed run, and no repair changed a classical algorithm. Appendix~\ref{app:selection} records which claims ran, when missing models were retrained, and how unstated parameters were set.

The census follows the same ladder, adapted to each article's own quality criterion and with its classical algorithm kept fixed (Fig.~\ref{fig:procedure}). It has three levels. At L0 we check that the released code runs; when no trained model was released, only the classical code is run. At L1 we repeat the article's own comparison on our workstation, as in the rung $S_1$. At L2 we change only the classical solver or device and compare at the article's quality measure, which combines the rungs $S_2$ and $S_3$ (Table~\ref{tab:census}). For a convolutional neural network (CNN) that estimates thermal conductivity, for example, L1 timed the article's own classical solver, and L2 replaced it with a direct solve of the same equations that was at least as accurate as the CNN \citep{audP00610}.

For two census methods, the phase-field CNN and the one-shot generator, L2 keeps the L1 comparison: the first's classical code already used a fast direct solver, and the second's got no faster in our test (Table~\ref{tab:census}; Appendix~\ref{app:selection}).

\subsection{Search, screening, and sample}\label{sec:search}
The article audit asks what a published comparison lets a reader verify. We searched four sources for articles that compare learning-based and classical methods in structural or thermal design, screened them with a language model, and drew a seeded random sample of the eligible records. Appendix~\ref{app:search} gives the search and eligibility rules.

\begin{figure}[htbp]\centering\includegraphics[width=\linewidth]{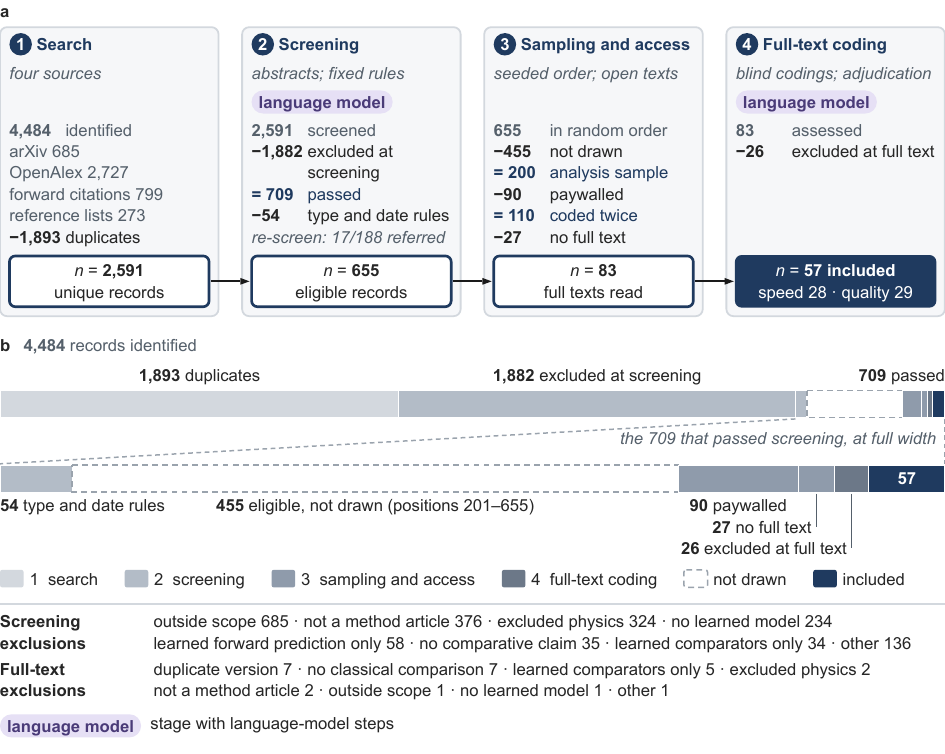}
\caption{Flow of records through language-model screening, seeded sampling, access assessment, two full-text codings, and adjudication. (a) Each stage lists the records it received, removed ($-$), and kept ($=$), and ends in its output; tagged stages include language-model steps. (b) Where all \N{N001} identified records left the flow, with the \N{N003} that passed screening shown at full width; the dashed segment holds eligible records outside the analysis sample. Blind re-screen referrals arose from unclear abstracts, not confirmed screening errors}\label{fig:flow}\end{figure}

Figure~\ref{fig:flow} follows the records from search to coding. Among the first 200 sampled records, those that were paywalled or had no readable full text were skipped, and some were excluded after reading. The coded set contains \N{N086} articles: \N{N087} whose main comparison concerns speed or cost (the speed claims) and \N{N151} whose main comparison concerns design quality only. The audit estimates therefore describe openly accessible records in this draw, not all publications.

\subsection{Primary outcome and rules}\label{sec:rules-summary}
Each article enters one outcome set, decided by its primary comparison, which a fixed rule selects (Appendix~\ref{app:coding}). Speed or cost comparisons form the speed-claim set, and quality-only comparisons stay separate.

Four pre-registered rules, adapted from the equal-quality and efficient-baseline rules of \citet{mcgreivy2024}, score each comparison (Appendix~\ref{app:rules}, Table~\ref{tab:rules}). R1 asks whether designs have matched quality or are compared at equal runtime or finite-element evaluation budget. R2 asks whether stated or measured evidence shows an inefficient classical baseline. R3 asks whether data-generation and training costs are reported with a break-even count or explicit per-query scope. R4 asks whether returned designs are verified against the physical model, with constraints and failures reported.

We used three verdicts. Fail required specific evidence of a violation. Verdicts score whether the reported comparison meets each rule; they do not judge an article's contribution, its correctness, or its authors' intent. Partial marked incomplete evidence or a likely concern. Pass meant that the rule was met or no violation could be shown; for R2, an unknown solver therefore passed. A pass is not evidence that a classical implementation is efficient. The headline counts only failures; counting partial verdicts as failures is a sensitivity analysis. R3 does not apply to methods that learn per problem without offline training.

\subsection{Coding, adjudication, and research software}\label{sec:coding}
Two passes by Claude-family language models, blind to each other, coded each accessible full text with the same instructions and recorded a source location for every judgment. Claude Opus 5.5 resolved their disagreements from the source text after agreement was measured. Because coders and adjudicator came from one provider, their errors may be correlated. A blind check by models from another provider and an unblinded authors' review, which is not independent validation, also examined the coding (Appendix~\ref{app:coding}).

The authors designed the study and wrote the manuscript. Claude assisted in developing the research code and organizing the evidence files, and Claude and OpenAI GPT models assisted in reviewing and editing the draft. The authors inspected the data, code, and results and take full responsibility for the content.

\subsection{Deviations and statistics}
Some choices were made after the protocol was fixed, and we call them post hoc. In the audit, model passes with model adjudication replaced the planned human verification. Appendix~\ref{app:deviations} lists the other post hoc choices: four for OAT, made after its re-test runs, and three later additions.

Audit rates are reported with exact Clopper--Pearson 95\,\% confidence intervals (CIs), which describe uncertainty in a proportion \citep{clopper1934}. These intervals cover the sampling of the coded judgments only, not screening, access, or coding uncertainty. Appendix~\ref{app:coding} gives the interval and agreement calculations.

Each OAT comparison uses five problems fixed in advance. For fixed budgets, we divide the median classical time by the median time of the learning-based method. For matched quality, we take the median of the five per-problem ratios.

If no tested budget matches within 150 iterations, the problem is censored: its time to quality was not observed. We rank these problems after all matched problems when calculating the median. For five problems, the median is the third ratio in ascending order. If fewer than three problems match, there is no observed median speed-up.

\section{Results}\label{sec:results}
\subsection{Implementation changes outweigh moving the released code to our machine}\label{sec:implementation-results}
\begin{figure}[htbp]\centering\includegraphics[width=\linewidth]{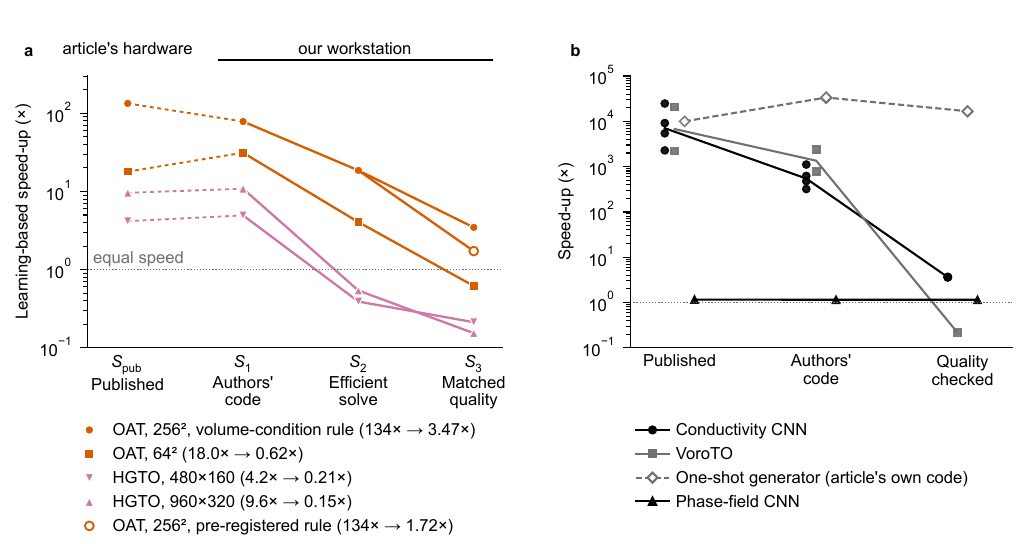}
\caption{Speed-up at the comparison levels of Fig.~\ref{fig:procedure}: published, authors' code, efficient solve, and matched quality. (a) Optimize Any Topology (OAT), a diffusion model, and HGTO, a graph-based method; the dashed segment is the value implied by the published timings on the article's hardware. At matched quality, the filled $256^2$ OAT marker uses the post hoc volume-condition rule and the open one the pre-registered rule. (b) Census comparisons with article-specific quality checks; open one-shot markers are L1 against its own code. Phase-field CNN compares convergence times, with lower learned compliance and an unchanged classical solver. Values below one favor classical optimization. OAT matched-quality points use the first budget of our graphics processing unit (GPU) baseline that meets the rule, identified after the runs; evaluator time is not counted. The axis is logarithmic. CNN, convolutional neural network}\label{fig:decomp}\end{figure}

Improving the classical implementation changed the comparison more than moving the released code to our workstation. We tested released code for eight methods (Table~\ref{tab:census}): two in-depth re-tests, four census comparisons with evaluated quality, and two unresolved census cases (Fig.~\ref{fig:decomp}). The last group includes a classical-only run because no trained learned model was released. For OAT and HGTO, the speed-up implied by the article and the speed-up we measured with the same released codes differed by a factor no larger than \N{N093}.

In the four cases of Fig.~\ref{fig:decomp}a, OAT at both grid sizes and HGTO on its two largest meshes, our efficient solver for the equilibrium equations (the state solve) reduced the same-machine speed-ups \N{N402} while keeping the classical algorithm. Matching quality then changes the ratio again, as the next two subsections show. Appendix~\ref{app:comparison} gives the per-case reductions and the accounting identity that links the rungs of the ladder.
\begin{table}[htbp]\caption{Eight released-code tests and one stopped reconstruction check. Optimize Any Topology (OAT), a diffusion model, and HGTO, a graph-based method, were fixed before any timed run for in-depth re-tests; a third in-depth re-test stopped at our reconstruction check. The frozen census covers every speed claim in the analysis sample that links public code (\N{N300} of \N{N087}); code ran for \N{N301}, including one classical-only run, and the seventh is that stopped reconstruction, excluded from the eight-method total. Factors are classical time divided by the learning-based method's time. Reported factors are derived from the times each article reports on its own hardware; L1 repeats own-code comparisons here, and L2 changes only classical solve or device, with unstated parameters set by rule. For the one-shot generator, the tested threading change did not speed the classical solve, so L2 retains L1 at matched quality. The last column gives each quality criterion; phase-field CNN compares convergence times at lower learned compliance. CNN, convolutional neural network; GPU/CPU, graphics/central processing unit; SIMP--OC, solid isotropic material with penalization and an optimality-criteria update.}\label{tab:census}\centering\scriptsize
\begin{tabularx}{\linewidth}{@{}>{\raggedright\arraybackslash}p{0.16\linewidth}>{\raggedright\arraybackslash}p{0.16\linewidth}>{\raggedright\arraybackslash}X>{\raggedright\arraybackslash}p{0.22\linewidth}@{}}\toprule Method&Reported&Same-machine outcome&Quality criterion\\\midrule
\multicolumn{4}{@{}l}{\emph{In-depth re-tests}}\\
OAT diffusion generator, $256^2$ \citep{nobari2025}&\N{N016} (implied by published timings)&78.6$\times$ against OAT's released SIMP code; \N{N017} with an efficient GPU solve; at matched quality, single samples: \N{N101} under the pre-registered rule, \N{N018} with the volume condition&Compliance error from OAT's evaluator; the volume condition also limits material\\
HGTO graph-based reparameterization \citep{liu2026}&\N{N095} and \N{N096} (from the article's reported times)&\N{N068} and \N{N069} against the authors' SIMP--OC; the same algorithm with our GPU solve was faster on all six configurations at matched quality (HGTO took \N{N112} as long)&Time to convergence; final compliance from HGTO's independent evaluator\\
Halle et al. predictor \citep{halle2021}&259$\times$&Not compared: stopped at the pre-registered reconstruction check (\N{N077} within 1\,\%, 95 required)&Reconstruction check before timing\\\midrule
\multicolumn{4}{@{}l}{\emph{Census}}\\
Conductivity CNN \citep{audP00610}&2,280--24,747$\times$&320--1,112$\times$ at L1; 3.6$\times$ at L2 with more accurate solve&Effective conductivity against the published values\\
VoroTO \citep{audP00080}&2,200$\times$ and 20,775$\times$ (from projected classical times)&Fine-scale SIMP on GPU 4.7$\times$ faster on compliance (the 20,775$\times$ claim)&Compliance: the classical run reaches VoroTO's structural objective\\
One-shot generator \citep{audP00183}&About $10^4\times$&1.6--1.7$\times10^4$ at matched quality against own code (L2=L1)&Compliance matched to the generator's designs\\
Phase-field CNN \citep{audP00663}&1.16$\times$&1.145$\times$ against own sparse-direct code (L2=L1)&Article's iteration counts and compliances reproduced\\
CNN de-homogenization \citep{audP00140}&5--10$\times$ (comparator time from another article)&Not comparable: primary comparator not public; fine-scale quality not reconstructed&Not possible\\
Three-dimensional unit cell \citep{audP00647}&2,400$\times$&Not comparable: no released model; we stopped our one-core run of the released classical code before convergence, after \N{N319} (article: 2.4 h)&Not possible\\\bottomrule
\end{tabularx}\end{table}

\subsection{OAT retains a small, resolution-dependent advantage}\label{sec:oat-results}
\begin{figure}[htbp]\centering\includegraphics[width=\linewidth]{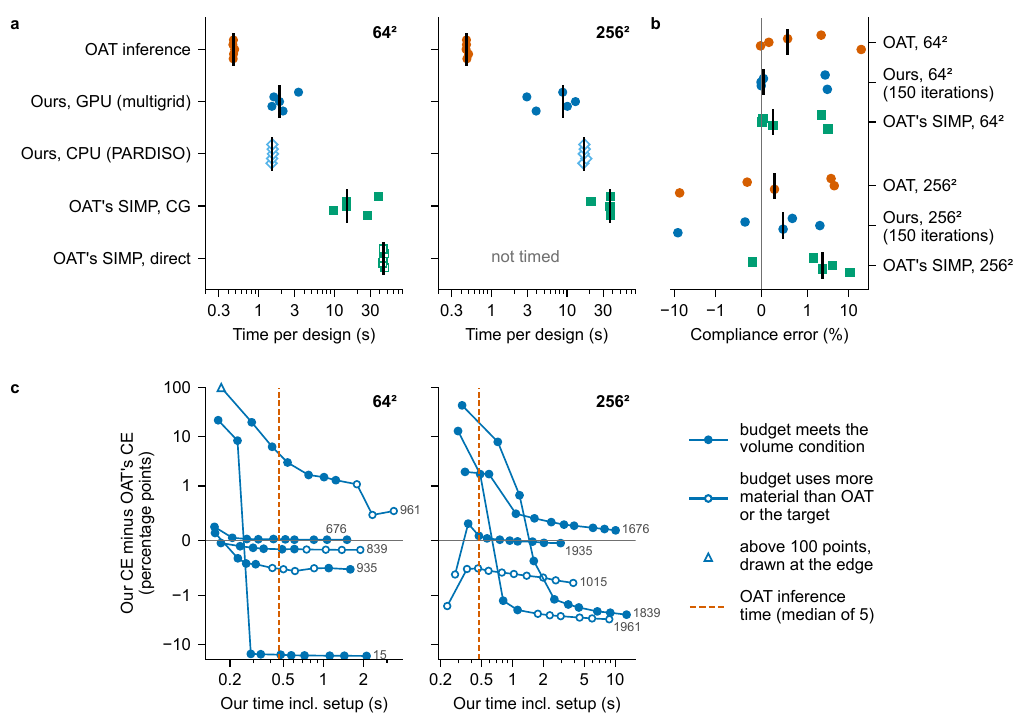}
\caption{Optimize Any Topology (OAT), a diffusion model: same-machine timing and quality. GPU/CPU, graphics/central processing unit; SIMP, solid isotropic material with penalization; CG, conjugate gradient; PARDISO, a sparse direct equation solver. (a) One-sample inference, setup-inclusive GPU and CPU classical runs, and authors' loop-only classical timing (OAT's direct-solver path was run at $64^2$ only); dots are problems and bars are medians. (b) Compliance error (CE) under OAT's evaluator. (c) Error of our GPU baseline minus OAT's error across fresh 5--150-iteration budgets, one curve per problem, numbered as in Online Resource 2 (three examples appear in Fig.~\ref{fig:designs}); filled marks at or below zero match and the dashed line is median OAT time. Error axes are signed linear-log, linear between $-1$ and $1$; values above 100 points in (c) are pinned to the edge}\label{fig:oat}\end{figure}

At $256^2$, one OAT sample took a median of \N{N042}, and our efficient classical run took \N{N062} for 150 iterations. Timed only to OAT's quality on the five primary problems, our classical run matched on \N{N102} under the pre-registered rule, which compares compliance error alone, and OAT was \N{N101} faster. Under the volume-condition rule, which also limits material, the classical run matched on \N{N052} and OAT was \N{N018} faster (Fig.~\ref{fig:oat}c). These advantages are far smaller than the \N{N016} implied by the published timings. The gap reflects the classical implementation and the quality match, not an error in those timings.

A larger set of 25 problems, added after the runs, gives a range. Across both rules and both problem sets, the single-sample advantage at $256^2$ ranged from \N{N405}.

\begin{figure}[htbp]\centering\includegraphics[width=\linewidth]{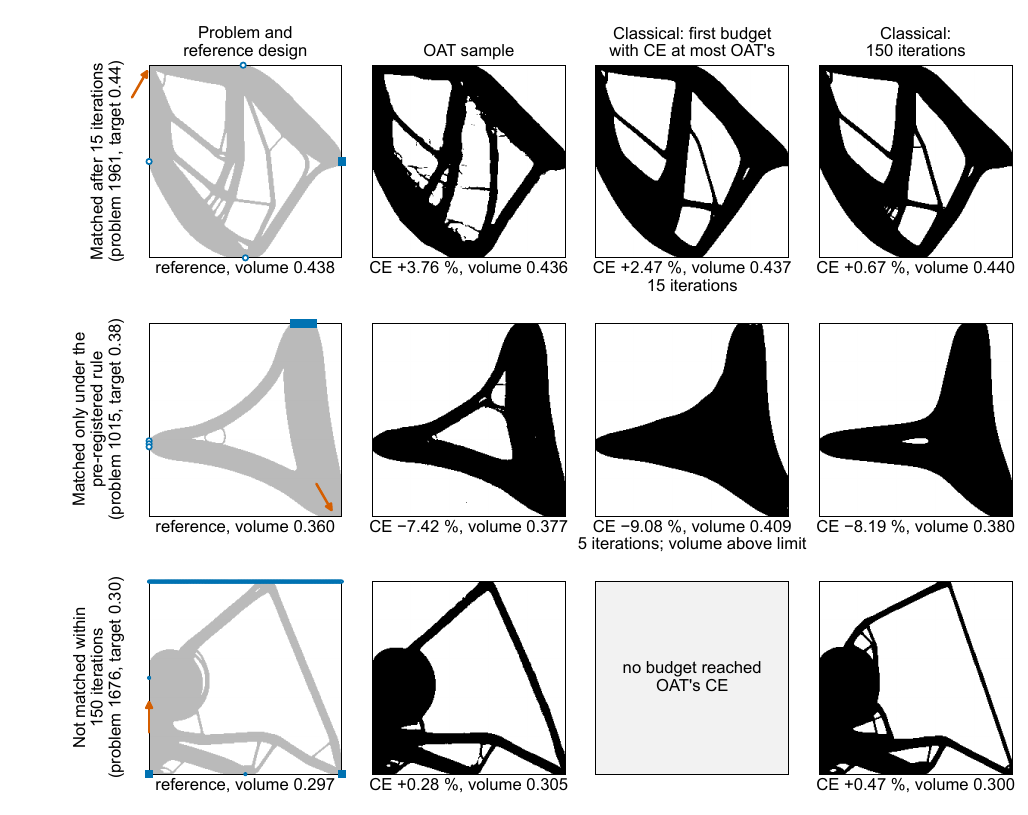}
\caption{Three $256^2$ test problems from the benchmark split of Optimize Any Topology (OAT), a diffusion model (rows), labeled by matching outcome. First column: the dataset's reference design (gray), against which compliance error is measured, with supports (filled squares fully fixed, open circles fixed in one direction) and the point load (arrow). Other columns: OAT's sample, the first classical budget whose compliance error is at most OAT's, and the classical design after 150 iterations. Designs are thresholded at 0.5 as OAT's evaluator scores them (black is solid); labels give compliance error (CE) and thresholded volume. A compliance-error match may not meet the volume-condition rule}\label{fig:designs}\end{figure}

Figure~\ref{fig:designs} shows why the matching rule matters. In one problem the classical design matched OAT after \N{N152} iterations. In a second it reached OAT's compliance error only by using more material, so it matched under the pre-registered rule but not under the volume condition. In a third it never matched within 150 iterations.

At $64^2$, the result depended on the classical solver. On the five primary problems, OAT was slower than the GPU baseline (\N{N039}) but faster than CPU PARDISO (\N{N105}) at matched quality. On the 25-problem GPU extension neither method was clearly faster (Appendix~\ref{app:oat-extension}). Because OAT's samples can exceed the target volume, the volume-conditioned comparisons score OAT's outputs as released; they do not time designs that respect the original volume target.

Drawing several OAT samples and keeping the best improves quality and can widen OAT's advantage. Picking the best sample, however, needs finite-element evaluations that our best-of-five sensitivity did not time, and the 150-iteration cap left some matches unmeasured. That sensitivity therefore picks the best sample as if its quality were known in advance and times inference only, so it is not an end-to-end deployment time.


\subsection{HGTO is slower at matched quality on our workstation}\label{sec:hgto-results}
Both HGTO outcomes, the time to convergence and the time to reach HGTO's final design quality, favor our classical implementation. HGTO's article reports run times for both methods; its stated contribution is comparable compliance at lower cost than a coordinate-based neural method, and the ratios here are derived from its reported times. On the primary outcome, time to convergence, our GPU port of the same SIMP--OC algorithm finished sooner in \N{N108} configurations; the exception was the inclined-load case, in which both classical implementations stopped at the update cap.

\begin{figure}[htbp]\centering\includegraphics[width=\linewidth]{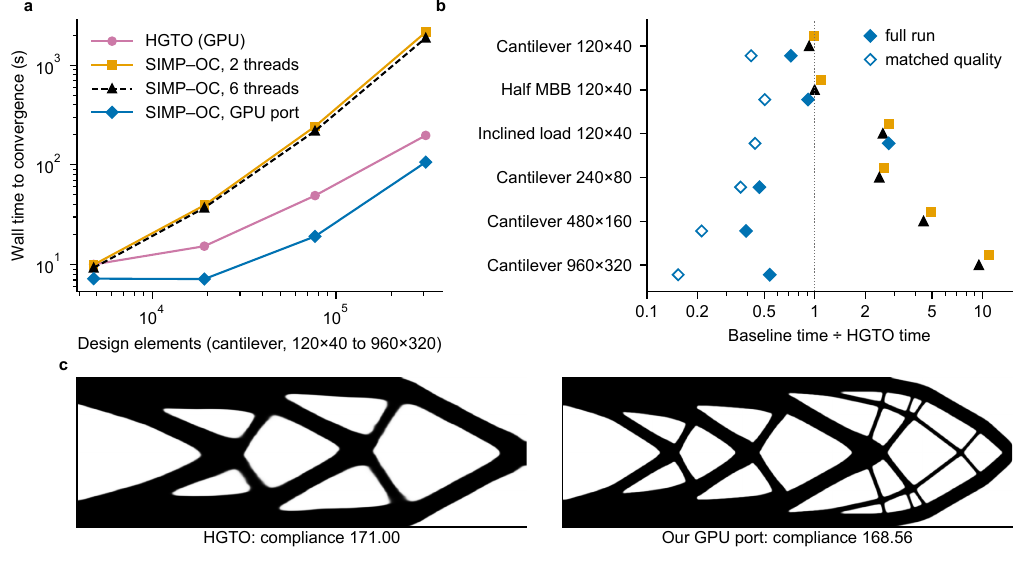}
\caption{HGTO, a graph-based neural design representation, and classical SIMP--OC on the same workstation. SIMP--OC combines solid isotropic material with penalization and an optimality-criteria update; GPU/CPU, graphics/central processing unit. (a) Wall time to convergence against mesh size for four cantilevers: HGTO on GPU, authors' SIMP--OC with two or six CPU threads, and our GPU port, as keyed by the plotted lines. (b) Classical time divided by HGTO time for six configurations (half MBB is the symmetric half of the Messerschmitt--B\"olkow--Blohm beam benchmark); values below one favor the classical baseline, and open diamonds show the GPU port at matched quality. (c) Final $960\times320$ designs and compliance under the authors' independent evaluator. Timing includes construction, optimization, and final evaluation; the inclined-load classical runs hit the update cap}\label{fig:hgto}\end{figure}

On the secondary outcome, time to HGTO's final compliance and volume, the GPU port was faster on all six configurations, and HGTO took \N{N112} as long. The largest mesh gave a \N{N075} classical advantage (Fig.~\ref{fig:hgto}). This reversal holds for the tested linear two-dimensional cases and our implementation; it does not show an error in the published timings.

Against the authors' own two-thread classical code on our machine, HGTO stayed ahead in five of six configurations (at cantilever $120\times40$ the two took about the same time) and was \N{N068} and \N{N069} faster on the two largest meshes. The reversal therefore comes from improving the classical solve, not from moving to our machine; Appendix~\ref{app:comparison} gives the timings.

\subsection{The code-linked census is mixed}\label{sec:census-results}
Tests of code for six public-code claims gave mixed outcomes (Table~\ref{tab:census}; Fig.~\ref{fig:decomp}b). For a convolutional neural network (CNN) that estimates thermal conductivity, the factor fell from \N{N303}, derived from the article's reported times, to 320--1,112$\times$ with the article's own classical code on one core of our machine, where hardware and stopping tolerance differ, and to \N{N305} against a direct solve of the same equations \citep{audP00610}. For VoroTO, a surrogate-based method for multiscale Voronoi structures, the fine-scale claim reversed: a GPU run of the fine-scale classical method reached the same objective \N{N309} faster \citep{audP00080}. A generator that returns a design in one shot and a phase-field convolutional network held their advantages against their articles' own classical codes \citep{audP00183,audP00663}, and two claims could not be compared at matched quality. Appendix~\ref{app:census} gives the timing and quality checks for each claim.

Read together, the comparable cases separate three situations. Where the published classical time came from a slower run or was never measured, the advantage shrank or reversed: our run of the conductivity CNN's own classical solver, and then a direct solve of the same equations, ran faster than the article's tabulated times, and VoroTO's fine-scale classical times were projections. Where the learned step worked inside a loop that already used an efficient solver, a modest advantage survived; the phase-field CNN reduced the number of iterations while the sparse direct solve stayed the same. 

Where an advantage survived only against the article's own algorithm, as for the one-shot generator, whose comparator used \N{N432} finite-element analyses, the census cannot say whether a more efficient classical algorithm would close the gap. Because the census never changes a classical algorithm, no held advantage shows that the article's code was as efficient as it could be, and these cases are single observations, not a pooled effect.

\subsection{Offline cost has a resolution-dependent break-even}\label{sec:breakeven}
\begin{figure}[htbp]\centering\includegraphics[width=\linewidth]{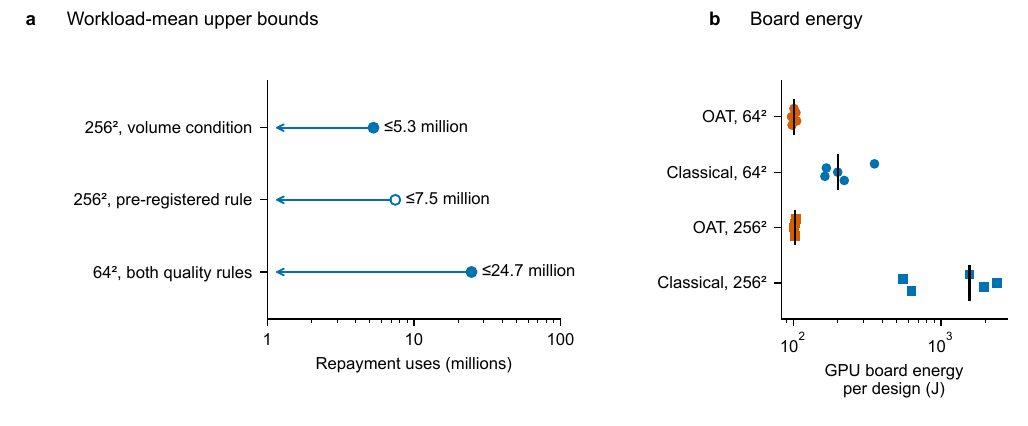}
\caption{Optimize Any Topology (OAT), a diffusion model: offline cost and graphics processing unit (GPU) board energy. (a) Upper bounds on the uses needed to repay data-generation and training cost, calculated from the mean saving across the five-problem workload at the first tested matching iteration budgets. Arrows indicate upper bounds because unmatched classical runs are censored at 150 iterations. Filled marks use the post hoc volume-condition rule; the open mark uses the pre-registered compliance-only rule. H100 offline time is divided by RTX 5090 per-design savings; these are GPU-time bounds, not financial or energy break-even estimates. (b) RTX 5090 board energy per design; dots are individual problems and bars are medians. Central processing unit (CPU) package and matched-quality energy were not measured}\label{fig:break}\end{figure}

A learning-based method pays its offline cost, generating training data and training, before it returns any design. We ask how many design uses would repay that cost through per-design time savings. Figure~\ref{fig:break} shows the estimates for OAT, together with separate measurements of GPU board energy; Appendix~\ref{app:offline} gives the offline budget and the calculation.

To ask when offline cost is repaid, we use the mean time saved per design across the five problems. We subtract each problem's inference time from its matched-quality classical time, then average the savings. Dividing the combined offline cost by this mean saving gives the repayment count.

Some classical runs reached the iteration cap without matching OAT. Their capped times understate the classical time, so the mean saving is a lower bound and the repayment count an upper bound. For this workload, the bounds are at most \N{N501} uses at $256^2$ (\N{N502} under the pre-registered rule) and at most \N{N503} at $64^2$; for comparison, the OAT article's own break-even scenario gives \N{N021} uses. Appendix~\ref{app:offline} gives the calculation.

These bounds divide H100 offline time by RTX 5090 savings, so they compare GPU time across devices and are not financial or energy break-even estimates. At $64^2$, OAT took longer than the classical code on three of the five problems, but the two problems left unmatched at 150 iterations give large savings, so the mean saving is positive although the median saving is negative. Scenarios built from median inputs can therefore differ; Appendix~\ref{app:median-cost} gives them.

\subsection{The audited articles span methods, physics, and dimensions}\label{sec:audit-results}
The \N{N086} audited articles span families of learning-based methods, physical objectives, and dimensions (Table~\ref{tab:sample}). Whereas the re-tests measured implementation and quality effects directly, the audit asks how well published comparisons document the conditions needed to interpret a speed claim. These articles supply the exploratory judgments and the public-code census; the two in-depth methods lie outside this sample. Figure~\ref{fig:flow} shows the access and eligibility losses, with counts in Appendix~\ref{app:reporting}.

\subsection{Unmatched comparisons and an efficiency blind spot}
The audit found comparisons at unmatched quality, while the efficiency of the classical solver was often impossible to judge from the text. In the adjudicated verdicts, quality-matching rule R1 failed for \N{N005} of speed claims and baseline-efficiency rule R2 for none, so failures of either rule also number \N{N434} of \N{N087}. This R1-or-R2 count corresponds to the weak-baseline definition of \citet{mcgreivy2024}.

Our own later review, which saw the models' verdicts, raised the count of R1-or-R2 failures to \N{N435} of \N{N087}. The adjudicated count remains primary because it was fixed before that review; the difference measures coding uncertainty and validates neither rate.

The audit cannot establish how common inefficient classical solvers are. Rule R2 found no demonstrated failure, but missing implementation and timing details left efficiency undecidable for \N{N113} speed claims. An unnamed solver alone does not meet the failure criterion; Appendix~\ref{app:reporting} gives the reporting counts.

Offline cost and output verification also had reporting gaps: the full-cost rule R3 failed for \N{N089} applicable speed claims and the verified-output rule R4 for 4 of \N{N087} (Table~\ref{tab:audit}). These outcomes describe this accessible, model-coded set; Appendix~\ref{app:reporting} explains why its rates cannot be compared directly with the fluid-equation audit.

In \N{N115} failures of the quality-matching rule R1, learned inference was compared with a full classical run while the learned output was measurably worse.

In one coded example, the printed mean times imply a \N{N138} factor \citep{audP00176}. The network's mean compliance error was \N{N139}, computed only over the outputs that remained after \N{N140} of the generated outputs had been excluded as disconnected. Such factors can be interpreted only together with design quality and failure rates.
\begin{table}[htbp]\caption{Adjudicated fail and fail-or-partial rates. Fail requires evidence of a violation; partial marks incomplete evidence or a likely concern (Sect.~\ref{sec:rules-summary}). R3 is inapplicable to three speed claims; confidence intervals for these descriptive rates appear in Online Resource 4.}\label{tab:audit}\centering\small
\begin{tabular}{@{}llrr@{}}\toprule Stratum&Rule&Fail&Fail or partial\\\midrule
Speed&R1 matched quality/time&8/28 (29\,\%)&22/28 (79\,\%)\\
Speed&R2 efficient baseline&0/28 (0\,\%)&11/28 (39\,\%)\\
Speed&R3 full cost&7/25 (28\,\%)&20/25 (80\,\%)\\
Speed&R4 verified output&4/28 (14\,\%)&20/28 (71\,\%)\\
Speed&R1 or R2&8/28 (29\,\%)&24/28 (86\,\%)\\
Speed&Any rule&13/28 (46\,\%)&27/28 (96\,\%)\\
Quality&R1&1/29 (3\,\%)&27/29 (93\,\%)\\
Quality&R4&5/29 (17\,\%)&26/29 (90\,\%)\\\bottomrule\end{tabular}\end{table}

\subsection{Reporting and agreement limit the audit}\label{sec:agreement}
Missing information and coder disagreement limit what these counts mean. Most included articles did not provide the code, data, and timing information needed to reconstruct their comparisons: \N{N032} linked code, and \N{N120} speed claims stated a full timing protocol (Fig.~\ref{fig:report}; indicators in Appendix~\ref{app:reporting}).

We measured agreement between the two coding models with Cohen's $\kappa$, which is 0 at chance-level agreement and 1 at perfect agreement \citep{cohen1960}. The coders disagreed most on quality matching (R1, \N{N035}) and baseline efficiency (R2, \N{N036}; Table~\ref{tab:agreement}), the two rules behind the headline count. Appendix~\ref{app:coding} gives all agreement values and the changes made in review.

The cross-provider check of Sect.~\ref{sec:coding}, with models from another provider, did not resolve this uncertainty. Their agreement with the original coders was of the same order as the original coders' agreement with each other, and R1 agreement stayed low for every pair of coders (\N{N522}).

Coders also disagreed on which articles are speed claims, so each coder's failure rate rests on a different set of articles. These rates describe coder-dependent samples, not alternative judgments on one fixed set (Appendix~\ref{app:coding}).
\begin{figure}[htbp]\centering\includegraphics[width=0.55\textwidth]{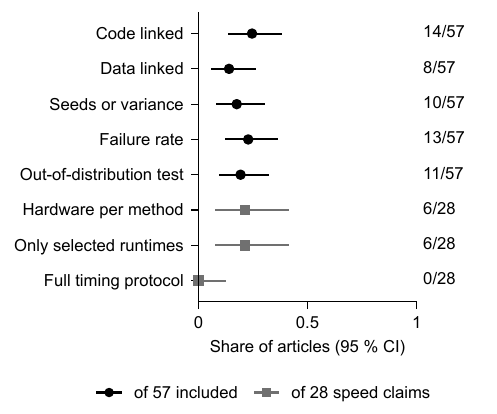}
\caption{Model-coded availability and reporting indicators, with exact Clopper--Pearson 95\,\% confidence intervals for proportions \citep{clopper1934}. Code, data, seeds, failures, and out-of-distribution testing cover the speed-claim and quality-only groups; hardware and timing indicators cover speed claims. Shares describe reporting in this coded set}\label{fig:report}\end{figure}

\section{Discussion}\label{sec:discussion}
\subsection{What the controlled re-tests establish}
The controlled re-tests show how much the classical implementation alone changes a comparison. On one workstation, replacing only the classical equation solver, with the optimization update unchanged, cut the speed-ups of the two in-depth methods by up to \N{N530}. In these tests the implementation mattered far more than moving the released codes to our machine, which changed the ratios by at most a factor of \N{N093}.

Results across OAT, HGTO, and the public-code census were mixed, which limits the practical conclusion. At matched quality, OAT kept a \N{N405} single-sample advantage on $256^2$ grids; at $64^2$, the outcome depended on the classical solver. HGTO was slower at matched quality in every tested case, and two census claims held against their articles' own classical codes. Learning-based methods may also offer design diversity, interaction, differentiable priors, and exploration, which the audit's rules do not score.

The choice of classical device also changes the comparison. At $64^2$, our CPU direct solver (PARDISO) needed \N{N063} for 150 iterations against \N{N064} for the GPU solver. At the shorter budgets matching OAT's quality, PARDISO's setup made it slower than the GPU solver. The faster solver therefore depends on the budget; we report both.

Offline cost can be repaid only if the mean time saving per design across the workload is positive; individual designs can still be slower. For our OAT workload, repayment needs at most \N{N501} uses at $256^2$, counted in GPU time across different GPUs. OAT's authors note that such accounting omits efficient parallel data generation and the value of access for users without large compute \citep{nobari2025}.

\subsection{Relation to prior work}
The results are consistent with evidence that neural reparameterization affects optimization outcomes \citep{herrmann2024,sanu2025}. The fluid-equation audit \citep{mcgreivy2024} found weak baselines in \N{N055} of its included articles, but that share cannot be compared with ours because domain, sampling, and efficiency judgments differ.

\subsection{Use beyond topology optimization}
The comparison ladder in Fig.~\ref{fig:procedure} can also organize speed evaluations of learned solvers for partial differential equations (PDEs) and simulation surrogates. It separates a published ratio from a local run with the released classical code, then tests implementation and quality under stated conditions. The ratios between neighboring rungs (Fig.~\ref{fig:procedure}; Eq.~(\ref{eq:decomp})) record the change due to execution context, implementation, and quality. They describe the tested comparisons and do not predict performance on another problem. The context ratio covers every difference between the published and the local runs, not only hardware. Applying the procedure requires a common quality criterion and compatible timing boundaries. \citet{nooraiepour2026} likewise emphasize solution verification and what each method can certify when selecting PDE methods. Our measured reductions remain specific to the tested problems and implementations; the comparison procedure can be reused wherever those evaluation conditions can be met.

\subsection{What a text audit can and cannot establish}
The text audit assesses whether articles report matched design quality, complete cost, and output verification. Quality-matching rule R1 tests whether the timed designs have comparable evaluated quality; baseline-efficiency rule R2 tests for evidence of an inefficient classical comparator. R2 fails a baseline only on stated or measured evidence of inefficiency, so an unknown solver passes. R2 produced no failures here. The R1-or-R2 count therefore measures only comparisons at unmatched quality, not how common inefficient classical solvers are, and the audit makes no judgment about any author's intent. The counts also depend on how the models coded the articles and on which articles were accessible; Sect.~\ref{sec:limitations} discusses both.

\subsection{Limitations}\label{sec:limitations}
\paragraph{Experimental scope.}
The re-tests are narrow: two in-depth methods, one workstation, and linear two-dimensional problems. A third in-depth re-test, of the predictor of \citet{halle2021}, stopped before timing because we could not reconstruct its released benchmark reliably. The census's only three-dimensional claim could not be compared, although \N{N155} speed claims in the coded set involve three dimensions. Timing also changes with device, software, batching, and scale, so differences from published figures describe our implementation and do not show that those figures are wrong.

\paragraph{OAT workload.}
The OAT comparison is small and partly post hoc (Sect.~\ref{sec:oat-rules}). OAT ran from its released code and checkpoints with guidance scale 1.0, its generation script's default, one sample per call, and without the batching and module compilation that the script applies or refinement, while our baseline was built for this GPU (Sect.~\ref{sec:classical}). The primary comparison uses five problems; the 25-problem extension, the volume-condition rule, the choice of the GPU baseline, and the best-of-five analysis were chosen after the runs, and the extension's tests were not corrected for multiplicity. The best-of-five analysis also left the finite-element evaluations that pick the best sample untimed, so it does not measure the complete time to produce a usable design.

\paragraph{Census interpretation.}
The census cannot detect a classical algorithm that is inefficient by design. Because it keeps each article's classical algorithm, it cannot tell whether a classical algorithm uses more finite-element analyses than an alternative would; this applies to every census row and to the audit's R2 verdicts. Two learned models were retrained from released code and may differ from the authors' trained models. One census claim ran in a container without CPU pinning, and some runs overlapped (Online Resource 6).

\paragraph{Coding reliability.}
The coders may share errors. Both coders and the adjudicating model came from one provider, and \N{N133} records do not identify the exact model used for their first coding pass. Language models can outperform crowd workers on some text-annotation tasks \citep{gilardi2023}, but in systematic reviews their accuracy matched that of humans only in some tasks \citep{khraisha2024}.

Changing the provider did not remove disagreement. Agreement between coders from different providers was similar to agreement between the two original coders. Because each coder retrieved its own texts, this check compares whole coding workflows and cannot isolate a provider effect. R1 verdicts depended on the coder, and the adjudicated R1 rate lay near the low end of the coders' range. The authors' review saw the adjudicated judgments, so it is not blind human replication; no independent human coding exists.

\paragraph{Screening and access.}
The sample covers only openly accessible articles, screened from their abstracts by one model (Sect.~\ref{sec:search}; Fig.~\ref{fig:flow}), so the audit rates may not hold for paywalled venues. Records that a blind re-screen referred for a full-text read were read only in the post hoc cross-provider check. The first 200 records included \N{N128} paywalled and \N{N131} without readable text; among the paywalled were \N{N158} articles in Structural and Multidisciplinary Optimization and \N{N156} in Computer Methods in Applied Mechanics and Engineering. The exact confidence intervals reflect sampling alone and omit screening, access, and coding uncertainty.

\paragraph{Energy and cost.}
The energy and cost results of Sect.~\ref{sec:breakeven} are partial. The energy measurements cover only the GPU board, not the CPU, and the repayment counts divide H100 offline time by RTX 5090 savings, so they are GPU-time scenarios. No matched-quality energy or financial break-even was measured.

\subsection{Minimum report for a speed claim}\label{sec:minimum}
Table~\ref{tab:minimum} lists what a reader needs to reconstruct a comparison. It specializes the general transparency and reproducibility goals of the REFORMS consensus checklist for machine-learning-based science \citep{kapoor2024reforms} for speed claims that compare learning-based and classical methods. It extends prior advice on equal-volume, equal-quality designs \citep{woldseth2022} with classical implementation, timing boundaries, and offline cost. A speed-up factor should be reported together with the objective and constraint values of both designs, including failed outputs. If quality differs, a curve of quality against time, or a comparison at a fixed time, is more informative than comparing inference time with the time of a converged classical run. State whether setup, loading, selection, evaluation, and refinement are timed, and give the hardware and use count behind any amortization claim.
\begin{table}[htbp]\caption{Minimum information for a reconstructable speed comparison. CPU/GPU, central/graphics processing unit.}\label{tab:minimum}\centering\small
\begin{tabularx}{\linewidth}{@{}p{0.24\linewidth}X@{}}\toprule Item&Report\\\midrule
Classical method&Formulation, update, filter, linear solver, and stopping rule\\
Hardware&CPU and GPU model, threads, and precision for each method\\
Timing&Setup, post-processing, warm-up, repeats, and statistic\\
Quality&Common finite-element objective, constraints, failures, and compared point\\
Offline cost&Data generation, training, hardware, and break-even use count\\
Artifacts&Code, weights, timing harness, and pinned environment\\\bottomrule
\end{tabularx}\end{table}

The reporting fields in Table~\ref{tab:minimum} can support evaluations of learned solvers and surrogates when their domain-specific objectives, constraints, and validation procedures are stated.

\FloatBarrier
\section{Conclusions}
An efficient classical implementation can greatly reduce a learning-based method's measured speed advantage without changing the classical optimization algorithm. We applied the comparison ladder to released code for eight methods on one workstation. For the two methods studied in depth, the diffusion model OAT and the graph-based HGTO, the efficient implementation alone cut the speed-ups by up to \N{N530}. We then timed the classical code until its design matched the learned design's quality; for OAT, this point was read off the completed runs, so it is not a stopping rule a user could apply. At matched quality, OAT was still \N{N405} faster when producing one design at a time on $256^2$ grids; at $64^2$, it was slower than our GPU solver and faster than our CPU direct solver. HGTO, by contrast, was slower at matched quality in every tested case. The six census cases add mixed evidence: two advantages held, one shrank, one reversed, and two comparisons remained unresolved.

The exploratory article audit classified \N{N434} of \N{N087} speed claims as compared without matched quality under model adjudication; an unblinded authors' review classified \N{N435}. These coder-dependent counts describe automatically screened records with retrievable full text. Classical solver efficiency was usually undecidable from article text. A reconstructable speed claim should identify the baseline implementation, compare on common hardware and stated design quality, and disclose the complete cost of producing a usable design.

The comparison ladder and minimum report in Table~\ref{tab:minimum} can guide authors and reviewers of other evaluations of learning-based against classical methods when the compared methods share an objective, constraints, and quality criterion. Offline cost also matters: OAT's primary workload at $256^2$ gives a GPU-time repayment bound of at most \N{N501} uses, not a financial or energy break-even.

Future work should extend the controlled re-tests to three-dimensional and nonlinear problems, and measure batched throughput and the complete time to produce a usable design. Independent, blinded human coding of an article subset would assess the reliability of the exploratory audit. Shared time-to-quality benchmarks could support both tasks by fixing quality criteria, recording failures, and timing validation and refinement.

\backmatter

\section*{Statements and Declarations}

\bmhead{Competing interests}

The search pool included records co-authored by the authors, and all were screened under the same protocol as every other record. Only one, \citet{drake2026}, entered the first 200 positions; it was excluded at full text because its heat-sink problem includes fluid flow, so no co-authored record entered an audit rate. The authors report no known financial interests.

\bmhead{Funding}

This work was supported by Santa Clara University.

\bmhead{Author contributions}

S. Yang: Conceptualization, Methodology, Software, Investigation, Formal analysis, Visualization, Writing -- original draft. J. Wang: Supervision, Conceptualization, Methodology, Writing -- review \& editing.

\bmhead{Acknowledgements}

The authors thank the Department of Mechanical Engineering at Santa Clara University for computational resources.

\bmhead{Data availability}

Coding records, evidence files, and figure data are deposited under the Creative Commons Attribution 4.0 International (CC BY 4.0) license at \url{https://github.com/nbbllxx0/ml-to-baseline-audit}. Online Resource 3 lists audited articles and DOIs; article texts are not redistributed.

\bmhead{Code availability}

Search, screening, coding, analysis, baseline, re-test, and census scripts are released under the MIT license in the same repository. Third-party code and weights must be obtained from their original sources at the recorded commits. OAT code is not redistributed because its repository has no license file.

\section*{Online Resources}
The Online Resources are not attached to this preprint; they are available from the repository named in the availability statements.

\noindent\textbf{Online Resource 1.} Audit protocol, queries, reason codes, coding instructions, and deviations.

\noindent\textbf{Online Resource 2.} Classical implementation, correctness gates, timing, energy, and re-test details.

\noindent\textbf{Online Resource 3.} Coded records, DOIs, verdicts, and locators with a machine-readable table.

\noindent\textbf{Online Resource 4.} Coding reliability, re-screen, access details, and the cross-provider check.

\noindent\textbf{Online Resource 5.} Stopped Halle et al. reconstruction.

\noindent\textbf{Online Resource 6.} Frozen census plan, repairs, deviations, and per-claim outcomes.

\section*{Replication of results}
Analysis scripts regenerate audit tables and figures from preserved coding and measurement files. The release includes the protocol, seeds, analysis code, timing harnesses, and pinned environment. Timed experiments used an RTX 5090 with CUDA 12.8 (13.0 for OAT's CuPy-based code) and performance-core pinning on a hybrid CPU. Census replication additionally requires MATLAB R2023b, MathWorks' numerical-computing software, a container for one claim, and third-party repositories at listed commits. Machine-dependent timings should be recomputed as same-machine comparisons.

\appendix
\renewcommand{\thetable}{A\arabic{table}}\setcounter{table}{0}

\section{Coded articles}\label{app:coded}
The speed stratum comprises \N{N087} articles \citep{audP00181,audP00176,audP00178,audP00183,audP00161,audP00149,halle2021,audP00740,audP00140,audP00550,audP00663,audP00549,audP00755,audP00890,audP01217,audP00080,audP00610,audP00647,audP00698,audP00705,audP00827,audP02187,audP00029,audP00786,audP00918,audP01517,audP00808,audP01974}. Online Resource 3 gives their DOIs and coding locators.

Tables~\ref{tab:speedarticles} and~\ref{tab:qualityarticles} list the included articles by first author, with the categories of Table~\ref{tab:sample} and the adjudicated verdicts.
\begin{table}[htbp]\caption{The \N{N087} speed claims. Factor: classical time divided by the learning-based method's time in the article's primary comparison, as coded, with factors above 100 rounded to integers; a dash means that the article gave no single factor. A dash under R3 means that the rule was inapplicable. Verdicts are the protocol-fixed adjudicated verdicts of Table~\ref{tab:audit} under the rules of Table~\ref{tab:rules}; they score the reported comparison, not the article's contribution. Online Resource 3 gives the evidence and locators for each verdict. Dim., dimension; 2D and 3D, two- and three-dimensional. Homogenization, architected-material homogenization; GAN, generative adversarial network; VAE, variational autoencoder.}\label{tab:speedarticles}\centering\scriptsize\setlength{\tabcolsep}{4pt}
\begin{tabularx}{\linewidth}{@{}>{\raggedright\arraybackslash}X>{\raggedright\arraybackslash}p{0.14\linewidth}>{\raggedright\arraybackslash}p{0.15\linewidth}l r llll@{}}\toprule Article&Method&Physics&Dim.&Factor&R1&R2&R3&R4\\\midrule
\citet{audP00610}&Convolutional network&Thermal&2D&2{,}280$\times$&Pass&Partial&Pass&Pass\\
\citet{audP01974}&Language-model agent&Compliance&2D, 3D&3.33$\times$&\textbf{Fail}&Partial&--&Partial\\
\citet{audP00181}&Convolutional network&Compliance&3D&1.67$\times$&Partial&Pass&\textbf{Fail}&\textbf{Fail}\\
\citet{audP00827}&Surrogate&Stress&2D&10{,}000$\times$&Partial&Pass&Partial&Pass\\
\citet{audP00890}&Other&Compliance&2D, 3D&12$\times$&Partial&Pass&Partial&Partial\\
\citet{audP00183}&Other&Compliance&2D&10{,}000$\times$&\textbf{Fail}&Partial&Partial&Pass\\
\citet{audP00663}&Convolutional network&Compliance&2D&1.16$\times$&Pass&Pass&\textbf{Fail}&Partial\\
\citet{audP00140}&Convolutional network&Compliance&2D&5$\times$&Partial&Pass&Partial&Partial\\
\citet{audP02187}&Other&Compliance&3D&4$\times$&Partial&Partial&Partial&Pass\\
\citet{halle2021}&Convolutional network&Compliance&2D&259$\times$&Partial&Pass&Pass&Partial\\
\citet{audP00740}&Other&Compliance&2D, 3D&1.86$\times$&Pass&Pass&Pass&Pass\\
\citet{audP00698}&Convolutional network&Compliance&2D&--&Partial&Pass&Partial&Partial\\
\citet{audP00549}&Convolutional network&Other&3D&18{,}000$\times$&Partial&Partial&Partial&Partial\\
\citet{audP01517}&Other&Stress&2D&1.23$\times$&Pass&Partial&--&Pass\\
\citet{audP00786}&Surrogate&Homogenization&2D&500$\times$&Pass&Pass&Partial&Partial\\
\citet{audP00918}&Other&Compliance&2D&2.67$\times$&Partial&Pass&Partial&Partial\\
\citet{audP00161}&Convolutional network&Compliance&2D&5.75$\times$&\textbf{Fail}&Pass&\textbf{Fail}&Partial\\
\citet{audP00080}&Surrogate&Homogenization&2D&2{,}200$\times$&Partial&Partial&Pass&Partial\\
\citet{audP01217}&Convolutional network&Compliance&2D&10$\times$&\textbf{Fail}&Partial&Partial&Partial\\
\citet{audP00178}&GAN&Compliance&2D&240$\times$&\textbf{Fail}&Partial&\textbf{Fail}&\textbf{Fail}\\
\citet{audP00550}&Convolutional network&Compliance&3D&30$\times$&Partial&Pass&\textbf{Fail}&Partial\\
\citet{audP00029}&Convolutional network&Homogenization&2D&19$\times$&Partial&Pass&\textbf{Fail}&Partial\\
\citet{audP00705}&GAN&Compliance&2D&--&\textbf{Fail}&Partial&Partial&\textbf{Fail}\\
\citet{audP00647}&Convolutional network&Homogenization&3D&2{,}400$\times$&Partial&Partial&Pass&\textbf{Fail}\\
\citet{audP00755}&Surrogate&Homogenization&3D&2{,}160$\times$&\textbf{Fail}&Pass&Partial&Pass\\
\citet{audP00176}&Convolutional network&Compliance&2D&3{,}913$\times$&\textbf{Fail}&Pass&\textbf{Fail}&Pass\\
\citet{audP00808}&Neural reparameterization&Compliance&2D, 3D&1.885$\times$&Partial&Pass&--&Partial\\
\citet{audP00149}&Surrogate&Homogenization&3D&300{,}000$\times$&Pass&Pass&Partial&Partial\\
\bottomrule
\end{tabularx}\end{table}
\begin{table}[htbp]\caption{The \N{N151} quality-only articles. R2 and R3 are shown for speed claims only; Online Resource 4 gives R2 for three quality-only articles. Verdicts are the protocol-fixed adjudicated verdicts of Table~\ref{tab:audit} under the rules of Table~\ref{tab:rules}; they score the reported comparison, not the article's contribution. Online Resource 3 gives the evidence and locators for each verdict. Dim., dimension; 2D and 3D, two- and three-dimensional. Homogenization, architected-material homogenization; GAN, generative adversarial network; VAE, variational autoencoder.}\label{tab:qualityarticles}\centering\scriptsize
\begin{tabularx}{\linewidth}{@{}>{\raggedright\arraybackslash}X>{\raggedright\arraybackslash}p{0.16\linewidth}>{\raggedright\arraybackslash}p{0.16\linewidth}lll@{}}\toprule Article&Method&Physics&Dim.&R1&R4\\\midrule
\citet{audP00165}&Convolutional network&Compliance&2D&Partial&\textbf{Fail}\\
\citet{audP01320}&GAN&Compliance&2D&Partial&Partial\\
\citet{audP00141}&GAN&Compliance&2D&Partial&Partial\\
\citet{audP02074}&Transformer&Multiphysics&2D, 3D&Partial&\textbf{Fail}\\
\citet{audP00081}&Convolutional network&Compliance&2D&Partial&Partial\\
\citet{audP00676}&Neural reparameterization&Compliance&2D&Partial&Partial\\
\citet{audP00148}&Neural reparameterization&Compliance&2D&Partial&Partial\\
\citet{audP00842}&Neural reparameterization&Compliance&2D, 3D&Partial&Partial\\
\citet{audP02042}&Reinforcement learning&Multiphysics&2D&Partial&Partial\\
\citet{audP01813}&GAN&Compliance&2D&Partial&\textbf{Fail}\\
\citet{audP00375}&Diffusion or flow&Compliance&2D&Partial&Partial\\
\citet{audP00136}&VAE&Compliance&2D&Partial&Partial\\
\citet{audP00358}&Diffusion or flow&Compliance&3D&Partial&\textbf{Fail}\\
\citet{audP00527}&Other&Compliance&2D, 3D&Pass&Partial\\
\citet{audP00006}&Other&Compliance&2D, 3D&Partial&Partial\\
\citet{audP02361}&Surrogate&Compliance&3D&Partial&Pass\\
\citet{audP01960}&Reinforcement learning&Stress&2D&\textbf{Fail}&Partial\\
\citet{audP00193}&Diffusion or flow&Compliance&2D&Partial&Pass\\
\citet{audP00373}&Neural reparameterization&Multiphysics&2D&Partial&Partial\\
\citet{audP00003}&Convolutional network&Compliance&2D&Pass&Partial\\
\citet{audP00051}&Neural reparameterization&Compliance&2D, 3D&Partial&Partial\\
\citet{audP00019}&Neural reparameterization&Multiphysics&2D, 3D&Partial&Partial\\
\citet{audP01971}&Reinforcement learning&Compliance&2D&Partial&Partial\\
\citet{audP02476}&VAE&Homogenization&2D&Partial&Partial\\
\citet{audP01219}&Transformer&Stress&2D&Partial&Pass\\
\citet{audP00309}&VAE&Stress&2D&Partial&Partial\\
\citet{audP01232}&Convolutional network&Compliance&2D, 3D&Partial&\textbf{Fail}\\
\citet{audP00092}&Neural reparameterization&Compliance&2D&Partial&Partial\\
\citet{audP01105}&Surrogate&Homogenization&3D&Partial&Partial\\
\bottomrule
\end{tabularx}\end{table}
\section{Timing, sensitivity, and coding detail}\label{app:detail}
\setcounter{table}{0}\renewcommand{\thetable}{B\arabic{table}}
\subsection{Audit decision rules}\label{app:rules}
Table~\ref{tab:rules} lists the four decision rules; Online Resource 1 gives the full protocol and coding instructions.
\begin{table}[htbp]
\caption{Fixed rules. R1 checks matched quality or time, R2 classical efficiency, R3 full cost, and R4 output verification. The full protocol and examples are in Online Resource 1. An unknown condition scores pass under the conservative fail-only headline. GPU/CPU, graphics/central processing unit.}\label{tab:rules}
\small\begin{tabularx}{\linewidth}{@{}p{0.17\linewidth}X X@{}}\toprule
Rule & Fail & Partial\\\midrule
R1: matched quality or time & Inference against a converged classical run with worse learned output and no match; speed claim without quality comparison; or post-processing counted in quality but not in time & Loose matching or unstated stopping\\
R2: efficient baseline & Our re-implementation at least three times faster on the same hardware class and size; an abstract reports a weaker baseline than the results; unjustified GPU--CPU comparison at $\geq 10^5$ degrees of freedom; excessive iterations; or classical iteration at least ten times slower than a strong baseline & Suspected inefficiency without a comparable measure\\
R3: full cost & Neither data nor training cost given & Training cost without data-generation cost, or no amortization or per-query scope\\
R4: verified output & Only image-similarity metrics or surrogate predictions, or examples chosen without a stated rule & Finite-element verification without violations or failure rates\\\bottomrule
\end{tabularx}
\end{table}

\subsection{Additional reporting counts}\label{app:reporting}
The exploratory audit also found that both data and training cost were missing from \N{N089} applicable claims. The efficiency of the classical solver could not be determined from the text for \N{N113} claims. These reporting counts concern the coded set and are presented with the other rule outcomes in Table~\ref{tab:audit}.

The flow, styled after Preferred Reporting Items for Systematic Reviews and Meta-Analyses (PRISMA), in Fig.~\ref{fig:flow} reports identification through adjudicated coding \citep{page2021}. Of the first 200 sampled records, \N{N128} were paywalled and \N{N131} had no readable full text. After full-text exclusions, \N{N086} articles remained: \N{N087} speed claims and \N{N151} quality-only claims. They span families of learning-based methods, objectives, and dimensions (Table~\ref{tab:sample}). The primary claim favored the learning-based method in \N{N144} articles. 

A broader count describes failures across all four rules. At least one rule failed for \N{N009}. A separate sensitivity counts partial judgments as failures; this raised the R1-or-R2 share to \N{N012}. Table~\ref{tab:audit} gives the applicable denominators.

R2 failed for \N{N006} of speed claims, but the baselines' efficiency was undecidable from article text in \N{N113} cases and scored partial in \N{N114} (suspected inefficiency or unexplained hardware); only \N{N146} passed on stated evidence. Only \N{N119} named its linear solver, \N{N034} specified hardware for each method, and \N{N120} gave a full timing protocol. The rule deliberately does not fail an unnamed solver.

Cost and output reporting show further gaps. R3 failed for \N{N007} of applicable speed claims and R4 for \N{N008}. These model-coded, accessible-set shares cannot be compared directly with the fluid-equation audit's \N{N055} rate \citep{mcgreivy2024}. The fluid-equation audit could also fail a baseline from a method class much less efficient than the state of the art for the equation type, whereas R2 here requires stated or measured inefficiency.

Among the \N{N086} included articles, \N{N032} linked code, \N{N033} linked data, \N{N030} reported seeds or variance, \N{N029} a failure rate, and \N{N031} an out-of-distribution test.

Among \N{N087} speed claims, \N{N034} specified hardware per method, \N{N028} gave runtimes only for selected cases, and none gave the full timing protocol defined here (Fig.~\ref{fig:report}). These are model-coded reporting counts.

The quality mismatch is concrete in the failed R1 comparisons: \N{N115} involved learned inference against a full classical run while the learned output was measurably worse. In one coded example, the printed mean times imply a \N{N138} factor, while the network's mean compliance error was \N{N139}; the article reports that \N{N140} of outputs were disconnected and excludes them from those averages \citep{audP00176}. This illustrates why a time factor and the quality and failure denominators must be reported together.

\begin{table}[htbp]\caption{Selected categories of learning-based method and physics, with the complete dimension split, in the coded set; the method and physics rows do not sum to the total. Complete categories appear in Online Resource 3.}\label{tab:sample}\centering\small
\begin{tabular}{@{}lrrr@{}}\toprule Characteristic&Speed&Quality&All\\\midrule
Convolutional network&13&4&17\\Neural reparameterization&1&7&8\\Surrogate&5&2&7\\
Compliance physics&18&20&38\\Architected-material homogenization&6&2&8\\
Two-dimensional&17&19&36\\Three-dimensional&7&3&10\\Both&4&7&11\\\midrule Total&28&29&57\\\bottomrule\end{tabular}\end{table}

\subsection{Published speed-factor distribution}\label{app:published}
Among the articles in the coded set whose primary comparison concerns speed, \N{N015} gave a numeric speed-up, stated or derived by the coders from printed times, with median \N{N123}, nearest-rank interquartile range \N{N122}, and range \N{N014}. These factors follow the articles' own comparisons on different problems and computers, not a pooled performance estimate. The OAT article gives a break-even scenario of \N{N021} uses \citep{nobari2025}.

\subsection{Timing repetitions and variability}\label{app:timing}
Our fixed-budget solver timings and the HGTO runs used one discarded warm-up and three measured runs, OAT inference one warm-up and ten sampling runs per problem, and OAT's released classical code one warm-up and one timed run; medians are reported. The time-to-quality sweeps timed each budget once after a warm-up; in the pinned three-repeat runs of the same problems, the coefficient of variation was at most \N{N504} on the GPU and \N{N505} with PARDISO. Runs over 300 seconds in the neural reparameterization re-test and census runs over one hour were timed once.

The GPU solver was selected as the single strong OAT baseline after the runs. PARDISO is also reported because its \N{N063} fixed-iteration run is quicker than the GPU's \N{N064} at $64^2$. GPU board power was sampled every 100 ms and integrated over the timed window; CPU package energy and matched-quality energy were not measured.

\subsection{Protocol changes and numerical repair}\label{app:deviations}
The original audit protocol called for human verification. Before verification began, this became two blind-to-each-other model passes and model adjudication; the later authors' review was unblinded. A seeded random sample replaced an earlier pilot. After re-test runs, we included our setup in OAT design times, adopted the volume-condition rule, repeated timed stages with CPU pinning, and chose the GPU solver as the one strong OAT comparator. The 25-problem extension, the oracle best-of-five analysis, and the cross-provider check were post hoc. A rounding-level compliance derivative with the wrong sign, found in the extension, was clamped at zero; a repeated earlier run was unchanged to rounding precision. One three-dimensional census classical run was unconverged. Audit and baseline deviations are logged in Online Resource 1 and census deviations in Online Resource 6. The matched-quality runs were logged in advance to use the GPU solver; its choice as the one comparator and the best-of-five analysis are not logged there.

\subsection{OAT published comparator and execution checks}\label{app:oat-checks}
We use OAT's published GPU-SIMP timing row because our same-machine re-run times that code; the table's CPU SIMP row is nearly equal at $256^2$ (69.45 s against 68.30 s) but faster at $64^2$ (3.45 s against 9.13 s). 

OAT's released conjugate-gradient (CG) classical path, timed loop-only, took \N{N044}. Its multigrid path stopped with a fault in a sparse-library dependency, which a script without OAT code reproduces, so it was not timed (Online Resource 1).

\subsection{Exploratory OAT extension and sampling sensitivity}\label{app:oat-extension}
The post hoc extension uses 25 problems per size and five seeds. The separate best-of-five sensitivity selects the lowest-error sample using untimed finite-element evaluations, so it measures oracle selection rather than a timed selection procedure.

The post hoc 25-problem, five-seed extension gave \N{N172} and \N{N176} at $256^2$ under the volume-condition and pre-registered rules. Our baseline matched sooner on \N{N173} problems under the volume-condition rule (\N{N174}, exact two-sided sign test) and \N{N177} under the pre-registered rule (\N{N178}). At $64^2$ the factors were \N{N163} and \N{N167}; under the volume-condition rule our baseline matched sooner on \N{N164} (\N{N165}). These tests are exploratory and uncorrected. In the extension, \N{N533} OAT samples exceeded the original volume target, by at most \N{N534}, and the volume-condition rule allowed the classical design the same volume. Requiring the classical design to meet the original target instead left \N{N535} samples matched at $256^2$ and \N{N536} at $64^2$, with factors of \N{N537} and \N{N538}. These matches compare released outputs as scored; they do not time feasible designs at the original target.

 Selecting the best of five samples lowered compliance error but left the baseline matching only \N{N423} problems at $256^2$ under the volume-condition rule and \N{N428} at $64^2$ within 150 iterations. Only at $256^2$ under the pre-registered rule did the median problem match (\N{N425}). There, our time was \N{N424} OAT's sampling time, depending on whether five samples cost five calls or one. This is an oracle-selected, inference-only sensitivity; the evaluations selecting the winner were untimed. Because our runs stop at 150 iterations, selecting the best of several OAT samples may widen its advantage by an unmeasured amount.

The extension takes per-problem medians over five samples and uses exploratory, uncorrected two-sided sign tests, which count the direction of each paired time difference \citep{dixon1946}. No pre-registered check relied on a significance test.

\subsection{Median-input break-even scenarios}\label{app:median-cost}
OAT states \N{N021} uses, while its printed timing table yields \N{N022}. The paper does not list the inputs behind its stated count, so the two values may rest on different inputs; we report both without treating the difference as an error. For median inputs (the median time ratio applied to OAT's median inference time), our $256^2$ fixed-budget comparison yields \N{N023}, the matched-quality comparison with the volume-condition rule \N{N024}, and the pre-registered-only match \N{N103}. At $64^2$ the median inputs give no GPU-baseline saving at matched quality, and PARDISO yields \N{N404} uses.

\subsection{Coding membership and changed judgments}\label{app:coding}
Of the 28 adjudicated speed claims, first coding placed \N{N413} in the speed stratum and found an R1-or-R2 failure in \N{N414}; second coding placed \N{N415} and found \N{N416}. Adjudication gave \N{N010}, and the unblinded authors' review \N{N410}. In the authors' review, three R1 judgments changed from partial to fail, while the adjudicated verdicts keep them partial: two articles compared only voxel similarity to the optimized designs, not design quality \citep{audP00181,audP00647}, and one timed a surrogate output that was less accurate than the classical evaluation against it \citep{audP00827}; Online Resource 1 logs each change.

Coding provenance is incomplete. First-pass records name Claude Opus 5.5 for \N{N132}; \N{N133} do not preserve the exact model version. The second pass used Claude Sonnet 5, was blind to the first, and saw masked record identifiers. First-pass files were hashed before the second coding, and all codings before analysis.

Two later checks assessed how judgments varied. For the post hoc cross-provider check, GPT-6-Sol and GPT-6-Luna independently coded the same 110 records blind from the second pass's inputs, retrieving full texts themselves as every coder did (Online Resource 4). Separately, the authors reviewed every speed-stratum verdict with the adjudicated judgments visible and logged \N{N204} changes. Their review was unblinded and does not constitute independent validation.

The statistical summaries serve two distinct purposes: describing coded judgments and aggregating measured runtimes. Proportions receive exact Clopper--Pearson 95\,\% intervals \citep{clopper1934}. Cohen's $\kappa$ measures chance-corrected agreement between model codings, with 2,000-resample bootstrap intervals and raw agreement alongside it \citep{cohen1960}. Under the benchmarks of \citet{landis1977}, R1 and R2 agreement is fair. The audit intervals condition on coded verdicts and exclude screening, access, and coding uncertainty.

Coder disagreement limits what can be inferred from the audit rate. Agreement was lowest on quality matching (R1) and baseline efficiency (R2): \N{N035} and \N{N036}, compared with \N{N037} for full cost, \N{N038} for output verification, and \N{N079} for inclusion (Table~\ref{tab:agreement}).

The cross-provider check shows that changing the provider does not remove disagreement. Agreement between an OpenAI model and a Claude pass was of the same order as agreement within one provider, with lower values for GPT-6-Luna (Table~\ref{tab:agreement}). R1 agreement remained low for every coder pair (\N{N522}).

Failure rates also depend on which articles each coder assigns to the speed-claim set. GPT-6-Sol found an R1 failure in \N{N523} of its set and GPT-6-Luna in \N{N524}, compared with the adjudicated \N{N005}. These are different denominators, not alternative judgments on one fixed speed set. The rates therefore describe coder-dependent samples; even high authors-versus-adjudication agreement on R2 cannot establish solver efficiency where the necessary information is missing.

\begin{table}[htbp]\caption{Agreement between coders, as Cohen's $\kappa$ with a 95\,\% bootstrap confidence interval (CI); $\kappa$ corrects agreement for chance \citep{cohen1960}. Claude: first pass vs blind second pass, before model adjudication; these models were blind to one another but from one provider. OpenAI: GPT-6-Sol vs GPT-6-Luna; OpenAI vs Claude: range over the four pairs of one OpenAI model and one Claude pass; both from the post hoc cross-provider check (Online Resource 4). Authors: the later unblinded review against the adjudicated verdicts. $n$ and raw agreement refer to the Claude pair and the authors' review.}\label{tab:agreement}\centering\small\setlength{\tabcolsep}{4pt}
\begin{tabular}{@{}lrrrrr@{}}\toprule Item&$n$&Raw agreement&Claude $\kappa$&OpenAI $\kappa$&OpenAI vs Claude $\kappa$\\\midrule
Inclusion&110&88\,\%&0.62 (0.42--0.80)&\N{N510}&\N{N516}\\Stratum&110&70\,\%&0.53 (0.40--0.65)&\N{N511}&\N{N517}\\
R1&82&55\,\%&0.35 (0.21--0.49)&\N{N512}&\N{N518}\\R2&82&56\,\%&0.28 (0.12--0.43)&\N{N513}&\N{N519}\\
R3&82&83\,\%&0.66 (0.49--0.80)&\N{N514}&\N{N520}\\R4&82&70\,\%&0.56 (0.42--0.70)&\N{N515}&\N{N521}\\
R1, authors vs adjudication&28&89\,\%&0.83 (0.63--1.00)&--&--\\
R2, authors vs adjudication&28&93\,\%&0.85 (0.59--1.00)&--&--\\\bottomrule\end{tabular}\end{table}

We classified each article by its primary comparison so that speed and quality claims would not share a denominator. We selected the first quantitative comparison with a classical method in the abstract, conclusion, introduction, or main text, in that order; if none existed, we used the first qualitative comparison. A speed or cost primary outcome entered the speed stratum, while quality-only outcomes remained separate.

The article audit used two Claude-family language models to code accessible full texts under the same schema and instructions, with a source locator for each judgment. The first 200 sampled positions yielded 110 coding records; after access and eligibility losses, the included set contained \N{N086} articles. Inclusion and stratum agreement use all 110 paired records, whereas rule agreement uses the 82 included by both coders. The final adjudicated article set is different, so agreement and audit rates need not share a denominator. Claude Opus 5.5 resolved item-level disagreements by consulting the source text and recording reasons and locations; agreement was measured before this adjudication.

\subsection{Re-test selection and census detail}\label{app:selection}
Of \N{N086} included articles, \N{N087} made speed claims; \N{N300} linked public code and code ran for \N{N301}. Together with the two in-depth methods, these provide eight released-code tests, including one classical-only census run.

OAT was named in the protocol before coding because its published timings, offline-cost estimate, and checkpoints permit a controlled re-test. HGTO came from candidates flagged by coders when code or a trained model was public. It sat at position \N{N406} in the seeded order and had been coded in a pilot round before the random sample replaced that pilot. These selection histories make the cases illustrative rather than representative.

A third in-depth attempt stopped before timing because we could not reconstruct the released benchmark reliably. For the predictor of Halle et al., our classical port reproduced the stored compliance within 1\,\% for \N{N077} validation cases, below the pre-registered 95-of-100 gate \citep{halle2021}. The remaining mismatches are not yet explained, so this is a limitation of our reconstruction, not an assessment of the article's results (Online Resource 5).

The census separately covers every public-code speed claim in the audit sample. Before timed runs, we fixed its \N{N300} candidates; \N{N301} ran, while the seventh is the stopped reconstruction. Missing weights were retrained with the authors' code only when data generation and training fit within one GPU-hour; two models met that limit. Repairs were logged without algorithm changes, and unstated parameters such as filter radius follow a fixed rule (Online Resource 6).

Census level L0 checks that the authors' code runs. L1 repeats the article's comparison on our machine; L2 keeps the classical algorithm but changes the solve or device and matches the article's quality measure. These are the census counterparts of reproducing the comparison and then improving the classical implementation (Fig.~\ref{fig:procedure}); the algorithm itself remains fixed.

For two census methods, L2 keeps the L1 comparison (Table~\ref{tab:census}). The phase-field CNN of \citet{audP00663} speeds up a classical optimizer, which describes the material boundary as a smooth field, by letting a CNN predict some of its update steps. Its classical code already used a fast sparse direct solver, so we compare the time each version needs to converge and check the compliance of the final designs. The one-shot generator of \citet{audP00183} produces a design in a single network evaluation. Its classical comparator improves a design in several rounds, and we time it until the end of the first round whose design is at least as stiff as the generator's. The only L2 change we tested, running that code on more processor cores, made it slower, so L2 keeps the L1 result.

\subsection{Search and eligibility detail}\label{app:search}
The search combined arXiv title and abstract queries with OpenAlex title-and-abstract search, forward citations from six anchor articles, and three backward-citation sources \citep{priem2022}. Queries crossed topology optimization or architected-material terms with machine-learning terms; the search ended on 24 September 2026. Records were deduplicated by DOI, arXiv identifier, and normalized title. Eligible records were English journal papers, conference papers, or arXiv preprints from 2017 onward that compared a learning-based method for structural or thermal topology optimization or architected-material inverse design with a classical, non-learned method. Fluid, acoustic, photonic, and electromagnetic topology optimization, forward analysis alone, and machine-learning-only comparisons were excluded. Full queries, reason codes, and clarifications are in Online Resource 1.

A separate check examined unclear exclusions rather than assuming that they were screening errors. Separate Claude Sonnet agents, whose exact model version was not recorded, blindly re-screened a random 10\,\% of exclusions and referred \N{N066} for a full-text read. In the post hoc cross-provider check (Sect.~\ref{sec:coding}), two models from another provider found \N{N525} of these 17 records eligible at full text.

Claude language-model agents screened titles and abstracts, assigning one reason code per record; their model version was not recorded. Figure~\ref{fig:flow} follows the search's \N{N001} records through deduplication (\N{N002} unique records), screening (\N{N003}), and the mechanical type and date rules (\N{N004}). From this eligible set, we drew the analysis sample: records were sorted by identifier and permuted with seed 20260924, and the first 200 positions were selected. Full-text coding used only legally accessible open texts, including publisher-hosted open copies.

\subsection{Classical formulation and solver detail}\label{app:solver}
The modified solid-isotropic-material-with-penalization interpolation is
\begin{equation}
E_e=E_{\min}+\tilde\rho_e^p(E_0-E_{\min}),\qquad p=3 .
\label{eq:simp}
\end{equation}
The density filter \citep{bourdin2001} is
\begin{equation}
\tilde\rho_e=\frac{\sum_i w_{ei}\rho_i}{\sum_i w_{ei}},
\quad w_{ei}=\max\{0,r_{\min}-\|\mathbf x_e-\mathbf x_i\|\},
\label{eq:filter}
\end{equation}
and the optimality-criteria (OC) update with move limit $m=0.2$ and exponent $\eta=0.5$ is
\begin{equation}
\rho_e^{\rm new}=\max\{0,\rho_e-m,\min[1,\rho_e+m,\rho_e B_e^\eta]\},
\quad B_e=\frac{-\partial c/\partial\rho_e}{\lambda\,\partial V/\partial\rho_e}.
\label{eq:oc}
\end{equation}
The volume multiplier is found by bisection. Plane-stress bilinear elements use $E_0=1$, $E_{\min}=10^{-9}$, and $\nu=0.3$ unless the target requires otherwise. These choices define the controlled comparison: formulation, filter, penalty, update, and stopping rule remain fixed, while the state solve and device are the implementation changes.

The strong GPU state solver is matrix-free, geometric-multigrid-preconditioned conjugate gradient (CG) in double precision \citep{amir2014}, implemented in PyTorch \citep{paszke2019}. Sparse direct PARDISO provides a CPU alternative \citep{schenk2004}; SuperLU is the reference \citep{li2005}. On two standard problems at $64^2$ and $256^2$ elements, the GPU and PARDISO implementations passed pre-registered compliance and physical-density gates against SuperLU over 150 iterations before timing. Residual limits and per-case results are in Online Resource 2.

HGTO optimizes a graph-based neural representation rather than returning a diffusion sample. We re-ran its released code on six linear two-dimensional configurations \citep{liu2026} and timed it against the authors' two-thread and six-thread SIMP--OC implementations. We then replaced only the classical state solve with our GPU solver. The filter, projection, continuation, stage budgets, update, volume restoration, and termination stayed fixed; these are the original choices governing how the design is updated and stopped. Per-update and final-compliance gates passed before timing.

\subsection{OAT and HGTO comparison detail}\label{app:comparison}
OAT generates a design through a released diffusion model \citep{nobari2025}. We ran one bfloat16, 20-step denoising diffusion implicit model (DDIM) sample per call \citep{song2021ddim} without refinement on five prespecified test-split problems at each size. Its evaluator thresholds the raw density at 0.5, uses a direct state solve, and measures compliance error $\ce=(c-c_{\rm gt})/c_{\rm gt}\times100$ relative to the dataset design. Each classical budget from 5 to 150 iterations was a fresh setup-inclusive run.

OAT retains a speed advantage in the larger single-sample comparison, but its size depends on the classical implementation and the quality rule. Published timings imply \N{N016} at $256^2$ and \N{N019} at $64^2$ against OAT's GPU SIMP baseline \citep{nobari2025}. On our workstation, one unrefined sample took median \N{N042}; our setup-inclusive, 150-iteration GPU run took \N{N062} at $256^2$. At this fixed budget the factor was \N{N017} over the five primary problems (range \N{N106}), before quality was matched (Fig.~\ref{fig:oat}).

At $256^2$, under the pre-registered compliance-only rule, our baseline met OAT quality on \N{N102} primary problems within 150 iterations, and the median factor was \N{N101}. Under the volume-condition rule, it matched \N{N052} and the factor was \N{N018}. At $64^2$, \N{N053} matched and the volume-condition factor was \N{N039}; CPU PARDISO gave \N{N105}.

Under the volume-condition rule, every matched classical design also met the original volume target, so allowing OAT's volume changed no match; OAT's samples exceeded the target on two of five problems at each size, by up to \N{N506}.

OAT ran from its released code and checkpoints, one design per call, and without the batching and module compilation of its generation script, while our baseline was built for this GPU. Omitting refinement favors OAT's time but leaves its compliance error higher, making a classical match easier. Including our setup and applying the volume-condition rule favor OAT. At $64^2$, choosing the GPU solver favors OAT at 150 iterations but the classical run at matched quality.

Replacing the classical implementation while retaining its algorithm cut the same-machine speed-ups \N{N402} in the four cases of Fig.~\ref{fig:decomp}a: OAT at both sizes and HGTO at $480\times160$ and $960\times320$. The HGTO reduction grew with mesh size: \N{N531} at $240\times80$, versus \N{N532} on the three $120\times40$ configurations. For OAT at $256^2$, the comparison ladder gives the accounting identity
\begin{equation}S_3=S_{\rm pub}HIQ,\qquad
H=S_1/S_{\rm pub},\quad I=S_2/S_1,\quad Q=S_3/S_2.\label{eq:decomp}\end{equation}
For this OAT example, $H=\N{N407}$ is the change from published to local context, $I=\N{N408}$ is the implementation change, and $Q=\N{N409}$ is the quality-matching change. The context factor also includes any other difference between published and local setups. The identity summarizes the tested comparisons; it is not a predictive model.

The quality factor uses the primary five OAT problems and the volume-condition rule. OAT's classical code used its inexact CG path and excluded setup time. Both choices favor that code, making $I$ conservative.

The reversal follows from changing the classical implementation. Against the authors' two-thread SIMP--OC on our machine, HGTO remained \N{N068} and \N{N069} faster at $480\times160$ and $960\times320$ elements, near the separate-hardware factors \N{N095} and \N{N096} \citep{liu2026}. Six threads reduced but did not remove HGTO's advantage.

\subsection{Census outcomes}\label{app:census}
The census rows of Table~\ref{tab:census} cover all \N{N301} code-linked claims for which code ran; Fig.~\ref{fig:decomp}b shows the four with evaluated-quality comparisons. One conductivity convolutional neural network (CNN) factor fell from \N{N303}, derived from the article's reported times, to \N{N304} using the authors' code on our machine, then \N{N305} against a more accurate direct solve of the same equations \citep{audP00610}. The released classical solver itself ran \N{N306} times faster on our machine than the article's tabulated time, so that first reduction is not solely a machine effect: the hardware and solver tolerance differ, and its timing table does not state the CPU.

VoroTO, a Voronoi multiscale surrogate method, reversed against its fine-scale comparator: a GPU run of that single-scale SIMP, with a filter radius the article does not state, reached the article's structural objective \N{N309} faster \citep{audP00080}. This tests the 20,775$\times$ fine-scale claim, not the 2,200$\times$ homogenization claim, and compares compliance, not every potential benefit of multiscale design. VoroTO ran with a surrogate we retrained from its released notebooks, whereas the classical run used our GPU code.

Two factors held against the articles' own classical implementations. The one-shot generator remained \N{N313} faster at matched quality against its augmented-Lagrangian code, with no faster state solve found \citep{audP00183}. The designs of the generator as we retrained it from the released code nevertheless had \N{N314} times the converged compliance, and that classical code used \N{N432} finite-element analyses for \N{N433} degrees of freedom. Since the census keeps the algorithm, this row is L1 at matched quality and does not establish global classical efficiency.

The phase-field CNN factor remained \N{N316}, near the article's \N{N315}; our runs reproduced its iteration counts exactly and its compliances within 0.01, and its comparator was already a sparse direct solver \citep{audP00663}.

Two claims could not be compared at matched quality. For one CNN method, the primary least-squares comparator was not public, and we did not reconstruct the exact supports and fine-scale compliance evaluation from the released material \citep{audP00140}. The other released no trained model \citep{audP00647}. We stopped one run of its released classical code at iteration 253 of at most 600, after \N{N319} on one core here, against the 2.4 h the article reports; a six-thread projection came close to the article's figure. We therefore found no sign that its classical time is overstated. The census therefore has one fall, one reversal, two held, and two non-comparisons, plus the stopped reconstruction.

\subsection{Offline budget and repayment calculation}\label{app:offline}
OAT reports \N{N148} of data generation and \N{N149} H100-days of training \citep{nobari2025}. The workload-mean repayment bounds and board-energy measurements appear in Fig.~\ref{fig:break}. With combined offline cost $C_{\rm train}$, classical per-design cost $C_{\rm SIMP}$, and inference cost $C_{\rm inf}$, the use count that repays cost is
\begin{equation}\tau=C_{\rm train}/(C_{\rm SIMP}-C_{\rm inf}),\qquad C_{\rm SIMP}>C_{\rm inf}.\label{eq:break}\end{equation}

An unmatched problem has no measured time-to-quality. We know only that its classical time is at least the time of its capped, 150-iteration run. Using that capped time gives a lower bound on the saving, and therefore on the mean saving. Dividing offline cost by this lower bound gives an upper bound on the repayment count. With the first tested matching budget and untimed evaluation, the bounds are at most \N{N501} uses at $256^2$ (\N{N502} under the pre-registered rule) and at most \N{N503} at $64^2$.

Energy was measured separately. GPU board energy at $256^2$ was \N{N026} for OAT inference and \N{N025} for 150 classical iterations. Matched-quality and CPU energy were not measured.

\FloatBarrier

\bibliography{references}
\end{document}